\pdfoutput=1
\documentclass[a4paper,fleqn]{aa}
\usepackage{inputenc}
\usepackage{amsmath, amssymb}
\usepackage{graphicx}
\usepackage{subcaption}
\usepackage{multicol}
\usepackage{multirow}
\usepackage{booktabs}
\usepackage{natbib}
\bibpunct{(}{)}{;}{a}{}{,} 
\usepackage{hyperref}
\usepackage[varg]{txfonts}
\usepackage{enumitem}

\newcounter{refcount}
\newcommand{\prefcount}{\stepcounter{refcount}\therefcount}
\newcommand{\pprefcount}{(\stepcounter{refcount}\therefcount)~}

\newcommand{\HRD}{Hertzsprung–Russell diagram}
\newcommand{\deriv}[2]{\ensuremath{\frac{\partial {#1}}{\partial {#2}}}}

\newcommand{\fracdelta}[1]{\ensuremath{\frac{#1}{\sigma_{m_\lambda}^2}}}
\newcommand{\feh}{[\rm{Fe/H}]}

\newcommand{\teff}{\ensuremath{T_{\rm{eff}}}}

\newcommand{\sigmapi}{\ensuremath{\sigma_{\pi_0}}}
\newcommand{\threeparams}{$\teff, \log g$ and $\feh$}
\newcommand{\mags}{\ensuremath{(m_\lambda - M_\lambda - C_\lambda  A_K - \mu_d)}}
\newcommand{\dLdmu}{\ensuremath{\frac{\partial^2 L_{sed}}{\partial \mu_d^2}}}
\newcommand\MyHead[2]{%
\multicolumn{1}{l}{\parbox{#1}{\centering #2}}
}

\newcommand{\myimagesmall}[1]{\begin{center}\includegraphics[angle=0, width=0.48\textwidth]{images/#1}\end{center}}
\newcommand{\myimages}[2]{\begin{center}\includegraphics[angle=0, width=#2\textwidth]{images/#1}\end{center}}

\newcommand{\myimageTwo}[2]{\begin{center}\includegraphics[angle=0, width=0.45\textwidth]{images/#1}
    \includegraphics[angle=0, width=0.45\textwidth]{images/#2}\end{center}}

\newcommand\at[2]{\left.#1\right|_{#2}}

\addto\extrasenglish{%
\let\subsectionautorefname\sectionautorefname

}

\makeatletter
\renewcommand*\aa@pageof{, page \thepage{} of \pageref*{LastPage}}
\makeatother

\title{Isochrone fitting in the Gaia era.}
\titlerunning{Isochrone fitting in the Gaia era.}
\author{Alexey Mints\inst{\ref{inst1},~\ref{inst2}}\thanks{email: mints@mps.mpg.de} \and Saskia Hekker\inst{\ref{inst1},~\ref{inst2}}}
\authorrunning{A. Mints and S. Hekker}
\institute{Max Planck Institute for Solar System Research, Justus-von-Liebig-Weg 3, 37077 Göttingen, Germany \label{inst1} \and Stellar Astrophysics Centre, Department of Physics and Astronomy, Aarhus University, Ny Munkegade 120, DK-8000 Aarhus C, Denmark \label{inst2}}

\abstract{Currently galactic exploration is being revolutionized by a flow of new data: Gaia provides measurements of stellar distances and kinematics; growing numbers of spectroscopic surveys provide values of stellar atmospheric parameters and abundances of elements; and \textit{Kepler} and K2 missions provide asteroseismic information for an increasing number of stars.} 
{In this work we aim to determine stellar distances and ages using Gaia and spectrophotometric data in a consistent way. We estimate precisions of age and distance determinations with Gaia end-of-mission and TGAS parallax precisions.} 
{To this end we incorporated parallax and extinction data into the isochrone fitting method used in the Unified tool to estimate Distances, Ages, and Masses (UniDAM). We prepared datasets that allowed us to study the improvement of distance and age estimates with the inclusion of TGAS and Gaia end-of-mission parallax precisions in isochrone fitting.} 
{Using TGAS parallaxes in isochrone fitting we are able to reduce distance and age estimate uncertainties for TGAS stars for distances up to 1 kpc by more than one third, compared to results based only on spectrophotometric data. 
With Gaia end-of-mission parallaxes in isochrone fitting we will be able to further decrease our distance uncertainties by about a factor of 20 and age uncertainties by a factor of two for stars up to 10 kpc away from the Sun.
}
{We demonstrate that we will be able to improve our distance estimates for about one third of stars in spectroscopic surveys and to decrease  log(age) uncertainties by about a factor of two for over $80\%$ of stars as compared to the uncertainties obtained without parallax priors using Gaia end-of-mission parallaxes consistently with spectrophotometry in isochrone fitting .}


\date{XXX/YYY}

\begin{document}

\renewcommand{\figureautorefname}{Fig.} 
\renewcommand{\sectionautorefname}{Section} 
\renewcommand{\subsectionautorefname}{Section} 

\maketitle
\section{Introduction}
Understanding our Galaxy is essential to further our understanding of the Universe. We can learn how the Galaxy was formed and how it has evolved by studying its current structure. To this end, stellar spectroscopic surveys that cover many stars are essential. These surveys provide data on stellar kinematics, chemical compositions, temperatures and surface gravities. 
Another important ingredient for our understanding of galactic evolution are stellar ages \citep[see for instance][]{2016ApJ...831..139M, 2017A&A...602A..67A, 2017arXiv170600018M}. 
For single stars, measuring ages remains a challenge, as the age of the star is only weakly related to parameters that we can observe \citep{2010ARA&A..48..581S}. 
We can compare an observed star with a set of models of stars of different ages and chemical compositions. Physical parameters of the model (or a range of models) with spectroscopic parameters that are close to those of the observed star will provide estimates of the physical parameters of this star. There exists a variety of methods using this approach, generally labelled ``isochrone fitting''. The same method can be used to get absolute magnitudes of stars. These absolute magnitudes, combined with visible magnitudes from photometric surveys, can be used to derive values of distance and extinction in the direction of the star.

In \citet[hereafter Paper I]{Paper1} we presented a Unified tool to estimate Distances, Ages, and Masses (UniDAM) that uses isochrone fitting to estimate distances, ages and masses for stars from spectrophotometric data. This tool was applied to a set of publicly available spectroscopic surveys, resulting in a catalogue of distances, ages and masses for over 2.5 million stars. These results were released with the UniDAM source code\footnote{Uploaded to CDS and also available at \url{http://www2.mps.mpg.de/homes/mints/unidam.html}}. 

UniDAM is designed to be easily extendible to include new measurements. An important measure that was not included into UniDAM as provided in Paper I is parallax. In the current work, we introduce parallaxes into UniDAM and show the effect of parallax priors on the precision of age and distance estimates. For this we use parallax data from the Gaia catalogue. The released Gaia data (Gaia DR1, \cite{2016A&A...595A...4L}) contains parallaxes and proper motions for 2.5 million 
stars in its Tycho-Gaia astrometric solution (TGAS) sample \citep{2015A&A...574A.115M}. 
The upcoming Data Release 2 \citep[DR2, see][]{2017arXiv171010816K} will contain data with orders of magnitude increase in number of stars and in the precisions of the determined proper motions and parallaxes. Although far from being final, values of parallax and proper motions will have uncertainties not much higher than those predicted for end-of-mission performance.
These data will be complemented with data from Gaia radial velocity spectrometer \citep{2016A&A...585A..93R} and from ground-based spectroscopic surveys. This will include radial velocities and stellar physical parameters such as temperature, surface gravity and chemical composition for millions of stars and will provide an unprecedented view of the Galactic structure and kinematics.

As we show in this work, for a large fraction of stars in spectroscopic surveys distance will be almost entirely defined by parallax. Parallax does however not provide any other stellar parameters and spectroscopic (or photometric) measurements are required to derive physical properties of stars. In Paper I we have shown that distance modulus and age estimates are correlated. Therefore, it is important to estimate distance and age consistently, to avoid biases.

Here we present the results of incorporating Gaia parallax data into UniDAM\@. For other works that use Gaia data for isochrone fitting see for instance \citet{2017arXiv170704554M, 2018MNRAS.tmp..326Q}. 
Our work is novel in two ways. 
We demonstrate what precisions in distances and ages to expect with Gaia end-of-mission (EoM) parallax precisions. On top of that, for stars in the catalogue from Paper I that have TGAS counterparts, we publish updated distance and age estimates. 

If parallax priors are used in isochrone fitting, the impact on the distance estimates is straightforward.
For the large fraction of stars in the spectroscopic surveys the distance will be primarily defined by their parallax. However, the uncertainty $\sigma_d$ in distance $d$ obtained from the parallax increases with distance: approximately as $\sigma_d \approx \sigma_\pi d^2$, where $\sigma_\pi$ is the parallax uncertainty. The uncertainties in distances derived from spectrophotometric data are much less sensitive to the distance itself. Therefore, the contribution of spectrophotometric data becomes more relevant for more distant stars.

The impact of parallax priors on derived ages is not straightforward. Gaia parallaxes combined with visible magnitudes and extinction values constrain the absolute magnitudes in different photometric bands for a given star. This adds constraints on the models used to fit the star. If consistent with other data, these constraints reduce the range of physical parameters covered by fitting models and thereby may reduce the uncertainty of estimates of parameters.
As we show in this work, with TGAS data we improve the precision of age estimates for stars within 1 kpc from the Sun, as compared to results from Paper I. We also predict improvements in the precision of ages for stars with distances up to 10 kpc based on predicted parallax precisions of Gaia DR2 and future data releases. There is however a lower limit on the precision with which we can determine stellar age even in the case of the small parallax uncertainty that we expect from Gaia DR2. This is caused by the possibility that there is a range of models with different ages and spectroscopic parameters that fits into a narrow range of absolute magnitudes, dictated by parallax.

Along with this paper we publish an updated version of UniDAM\footnote{Available at \url{https://github.com/minzastro/unidam}} and the catalogue of updated distance, age and mass estimates for over $400,000$ stars from TGAS\@.

\section{Inclusion of parallax priors into UniDAM}
In this section we provide a brief description of the method applied in the original version of UniDAM (see Paper I for more details). We subsequently show how Gaia parallaxes can be incorporated in a consistent manner into UniDAM\@.

\subsection{Age and distance measurements without parallax}\label{sec:old_method}
UniDAM, presented in Paper I, utilises a Bayesian method of deriving stellar parameters from spectrophotometric data. By comparing observed stellar parameters (\threeparams) and visible infra-red magnitudes $m_\lambda$ to spectral parameters and absolute magnitudes $M_\lambda$ from PARSEC models \citep{PARSEC}, we derived probability density functions (PDFs) for log(age), mass, distance modulus $\mu_d$ and extinction in 2MASS K-band $A_K$ for a star. 

In Paper I we have shown that each model contributes a delta-function to the log(age) and mass PDFs for a given star. The contribution of each model to distance modulus and extinction PDFs has a bivariate Gaussian shape as the goodness-of-fit $L_{sed}$ of the spectral energy distribution is quadratic in $\mu_d$ and $A_K$:
\begin{equation}
L_{sed} = \sum_\lambda \frac{(m_\lambda - M_\lambda - C_\lambda A_K - \mu_d)^2}{2 \sigma_{m_\lambda}^2} - V_{corr}, \label{eq:old_lsed}
\end{equation}
where $m_\lambda$ is the observed visible magnitude with the corresponding uncertainty $\sigma_{m_\lambda}$; $M_\lambda$ is the absolute magnitude of the model, and $C_\lambda$ represents extinction coefficients. The summation is done over filters $\lambda$, for which photometry is available.
The last summand $V_{corr}$ is the volume correction, introduced to compensate for the fact that with a given field of view we probe a larger volume at larger distances. Volume correction is expressed as a natural logarithm of the square of the distance, which is in turn expressed here through distance modulus $\mu_d$:
\begin{equation}
V_{corr} = (0.4\mu_d+2) \ln 10.
\end{equation}
We refer the reader to Paper I for more discussion of volume correction and its effect on log(age) and distance modulus estimates.

The location of the Gaussian contribution of each model to the PDFs in distance modulus $\mu_d$ and extinction $A_K$ is calculated as optimal values (designated as $\mu'_d$ and $A'_K$) that minimise the goodness-of-fit $L_{sed}$ (\autoref{eq:old_lsed}) for this model.
The width of the above Gaussian in distance modulus is $\Delta_{\mu_d} = \sqrt{H^{-1}_{0,0}}$ and in extinction $\Delta_{A_K} = \sqrt{H^{-1}_{1,1}}$, where $H_{i,j}$ is the Hessian matrix:
\begin{equation}
H_{i,j} = \at{\frac{\partial^2 L_{sed}}{\partial x_i \partial x_j}}{\mu'_d, A'_K},\,\textrm{with} \,x = (\mu_d, A_K).
\label{eq:hessian}
\end{equation} 
Substituting \autoref{eq:old_lsed} into \autoref{eq:hessian} we obtain:
 \begin{equation}
    H = 
    \begin{vmatrix}
    \sum_\lambda\fracdelta{1}         & \sum_\lambda\fracdelta{C_\lambda} \\
    \sum_\lambda\fracdelta{C_\lambda} & \sum_\lambda\fracdelta{C_\lambda^2}
    \end{vmatrix}
. \label{eq:hessian_p1}
\end{equation}

An important property of $H$ and therefore, of $\Delta_{\mu_d}$ and $\Delta_{A_K}$ is that they depend exclusively on the photometric uncertainties $\sigma_{m_\lambda}$ and extinction coefficients $C_\lambda$. Hence, they are the same for all models of a given star.
Therefore, Gaussian components contributed by each model to the PDFs for a given star have exactly the same shape and differ only by location. Thus, it is possible to build PDFs in distance modulus and extinction by taking the distribution of $\mu'_d$ and $A'_K$ that is, the optimal values that minimise $L_{sed}$ for each model. These distributions should be smoothed with Gaussian kernels of width $\Delta_{\mu_d}$ for the distance modulus $\mu_d$ and $\Delta_{A_K}$ for extinction $A_K$, to account for the width of the Gaussians contributed by each model. This smoothing is, however, only a minor correction, because $\Delta_{\mu_d}$ and $\Delta_{A_K}$ are an order of magnitude smaller than the width of the distribution of $\mu'_d$ and $A'_K$, as shown in Paper I. 

The approach described above allows to produce PDFs in distance modulus and log(age) for stars for which spectrophotometric data are available. When Gaia parallaxes are available, this method has to be modified slightly. Below, we show how parallax data can be incorporated into UniDAM. 

\subsection{Age and distance measurements with parallax}\label{sec:method}
In the current work we aim to improve stellar parameters determined by UniDAM by incorporating Gaia parallax data, in addition to effective temperature $\teff$, surface gravity $\log g$, metallicity $\feh$ and visible magnitudes $m_\lambda$. This requires, as we show below, the use of an external value of the extinction.

Including parallax information into isochrone fitting is non-trivial. The transformation from parallax uncertainties to distance uncertainties is non-linear -- symmetric parallax uncertainties correspond to asymmetric distance uncertainties \citep{1998A&A...340L..35K, 2016ApJ...832..137A}. The asymmetry increases with increasing fractional parallax uncertainties. However, in the current work we show that asymmetries caused by transformation from parallax to distance uncertainties can be neglected in the majority of cases.

\subsubsection{The need for an extinction prior}\label{sec:extinction}
Constraints on stellar parameters can be degenerate for a star with a known parallax and unknown extinction. This is caused by the fact that the optimal values $\mu'_d$ and $A'_K$ which minimize $L_{sed}$ in \autoref{eq:old_lsed} are highly correlated. The aim of the parallax prior is to select models that give a matching $\mu'_d$ and remove other models from the consideration, thus reducing the uncertainties in the measured parameters. 
However, due to correlations between $\mu'_d$ and $A'_K$, this cannot be achieved without a prior on extinction. This can be illustrated with the following example. Let us assume that we have two models for which minimising \autoref{eq:old_lsed} (in other words, not using parallax data) results in optimal distance modulus and extinction values $\mu_1, A_1$ and $\mu_2, A_2$, with $\mu_1 \ne \mu_2$. Now let's assume that we know that $\mu = \mu_1$ from parallax measurement ($\mu_1 = -5 (\log_{10} \pi + 1)$, where $\pi$ is the measured parallax). If we use this without a prior on extinction we will still obtain two solutions $\mu_1, A_1$ and $\mu_1, A_2'$. Due to the correlation between $\mu'_d$ and $A'_K$, $L_{sed}(\mu_1, A_2')$ will not be much larger than $L_{sed}(\mu_2, A_2)$, and the contribution of the second model to PDFs in all parameters will not be removed. Effectively a difference in distance modulus between models will be shifted into a difference in extinctions. To avoid this degeneracy, we need to put a prior on extinction $A_K$. 

\subsubsection{Incorporating parallax and extinction into UniDAM}\label{sec:new_method}
To incorporate priors on parallax and extinction we modified the $L_{sed}$  goodness-of-fit in \autoref{eq:old_lsed}. We added two priors, namely a prior on parallax:
\begin{equation}
    Pr(\pi) = \frac{(\pi - \pi_0)^2}{2\sigmapi^2},
\end{equation}
where $\pi_0$ and $\sigmapi$ are measured parallax and its uncertainty.
For a prior on extinction we use the following expression:

\begin{equation}
    Pr(A_K) = \left\{  
    \begin{array}{l}
    \frac{(A_K - A_0)^2}{2\sigma_{A_0}^2}, \textrm{if}\,A_K > A_0, \\
    0, \textrm{otherwise}
    \end{array} \right.\label{eq:extprior}
\end{equation}
where $A_0$ and $\sigma_{A_0}$ are measured extinction in $K$-band and its uncertainty taken from an extinction map. Effectively this prior allows $A_K$ to vary between zero and $A_0$, and penalizes larger values. We chose this form because $A_0$ is the value of extinction at infinity, and we need to allow for lower extinctions for nearby stars. Three-dimensional extinction maps \citep[for example,][]{2018arXiv180103555G} can be used to get more realistic prior $Pr(A_K)$, however these maps do not yet cover all sky, so cannot be applied to all surveys. The prior will be updated in the future versions of UniDAM, when Gaia-based extinction maps will become available.

If we take parallax to be $\pi_0 \pm \sigmapi$ and extinction to be $A_0 \pm \sigma_{A_0}$, we can define the new goodness-of-fit for the model as a function of $\mu_d$, $A_K$ and parallax $\pi$, which in turn is a function of $\mu_d$:
\begin{align}
L_{sed}(\mu_d, A_K) =& \sum_\lambda \frac{(m_\lambda - M_\lambda - C_\lambda A_K - \mu_d)^2}{2 \sigma_{m_\lambda}^2} + \nonumber \\
&  Pr(A_K) + Pr(\pi) - V_{corr}. \label{eq:lsed}
\end{align}

Here, we add two quadratic terms for the extinction and parallax. This is done under the assumption that the uncertainties in these values have a normal distribution. Again, we can find optimal values of distance modulus $\mu'_d$ and extinction $A'_K$ that minimise $L_{sed}$. 
The method of finding $\mu'_d$ and $A'_K$ is given in \autoref{app:A}.
The term containing parallax $Pr(\pi)$ in \autoref{eq:lsed} can be expressed as a function of distance modulus $\mu_d$ as:
\begin{equation}
\frac{(\pi - \pi_0)^2}{2\sigmapi^2} = \frac{(10^{- 1 -0.2 \mu_d} - \pi_0)^2}{2\sigmapi^2}.
\end{equation}
Therefore, the re-defined $L_{sed}$ is no longer a quadratic function of $\mu_d$ and $H_{0,0} = \dLdmu$ is not constant for a given star. Instead, it is a function of the optimum parallax $\pi'$ for a given model:
\begin{equation}
H_{0,0} = \at{\dLdmu}{\mu'_d} = (0.2\ln 10)^{2} \frac{\pi'\left(2\pi' - \pi_0 \right) }{\sigmapi^2}+\sum_\lambda \frac{1}{\sigma_{m_\lambda}^2},
\label{eq:lsed_2nd_der}
\end{equation}
where $\pi' = 10^{-1 - 0.2 \mu'_d}$. The value of $H_{1,1}$ will also change to:
\begin{equation}
H_{1,1} = \left\{  
\begin{array}{l}
\sum_\lambda\fracdelta{C_\lambda^2} + \frac{1}{\sigma_{A_0}^2}, \textrm{if}\,A'_K > A_0, \\
0, \textrm{otherwise}
\end{array} \right..
\end{equation}
Thus $H_{1,1}$ remains constant for all models with $A'_K > A_0$ and zero for models with $A'_K \leq A_0$.
Thus, the assumption made in Paper I that the PDF in $\mu_d$ and $A_K$ is a Gaussian with the same width for each model for a given star is no longer valid.
Therefore, formally, we have to calculate PDFs in distance modulus and extinction for each model by directly evaluating \autoref{eq:lsed} over the two-dimensional grid. However, we argue that in many cases the contribution of each model to the PDF is still close to being Gaussian, and that $H_{0,0}$ is nearly constant, so we can keep the approach introduced in Paper I. To illustrate this, we consider two regimes:
\begin{itemize}
    \item ``photometry dominated'', where the fractional parallax uncertainty is large ($\sigmapi / \pi_0 \approx 1$). In this case the location of the minimum of $L_{sed}$ and the shape of $L_{sed}$ around this minimum are defined primarily by the photometric components. This is because the first summand in \autoref{eq:lsed} is much more sensitive to variations of $\mu_d$ than the third summand (parallax part). At the same time, the effect of the first summand in $H_{0,0}$ (see \autoref{eq:lsed_2nd_der}) is negligible, and $L_{sed}$ is nearly quadratic in $\mu_d$ and $A_K$. Thus, in case fractional parallax uncertainty is large, parallax can be considered as a minor correction to the method without the inclusion of parallaxes (see \autoref{sec:old_method});
    
    \item ``parallax dominated'', where the fractional parallax uncertainty is small ($\sigmapi / \pi_0 \ll 1$). In this case the location of the minimum of $L_{sed}$ is defined by the parallax component (see \autoref{eq:lsed}). 
    
    If the optimal parallax $\pi'$ derived for each model is not close to $\pi_0$, the parallax term in \autoref{eq:lsed} will be large, making the overall goodness-of-fit large, that is $L_{sed} \gg 1$. The contribution of the model to the PDF depends on $L_{sed}$: models with large $L_{sed}$ typically contribute little to the PDF if there are other models with much smaller $L_{sed}$. If there are no models with a much smaller $L_{sed}$, then the overall fit for the star under consideration is bad. Therefore, we can assume that optimal parallaxes $\pi'$ derived for each model are all very close to $\pi_0$.
    
    As long as the optimal parallax value $\pi'$ is close to $\pi_0$ we can approximate $\pi'$ by $\pi_0$ in \autoref{eq:lsed_2nd_der}, such that $H_{0,0}$ depends only on $\pi_0$ and $\sigma_{m_\lambda}$ and therefore, is a constant for all models for a given star. In this ``parallax dominated'' case we observed that $\Delta_{\mu_d}$ computed from the Hessian matrix $H$ is comparable to the scatter of optimal $\mu'_d$ for all models. Therefore, the impact of smoothing the PDF in distance modulus with Gaussian kernel of width $\Delta_{\mu_d}$ is no longer a minor correction and has to be accounted for. We can use the fact that for a given star, each model's contribution is close to a Gaussian, which shape is defined by $\sigma_{m_\lambda}$, $\pi_0$ and $\sigma_{\pi_0}$ and is the same for all models. Therefore, it is still valid to calculate the optimal $\mu'_d$ and $A'_K$ for each model and then smooth the resulting PDFs with $\Delta_{\mu_d}$ and $\Delta_{A_K}$.   
\end{itemize}

With the precision of spectrophotometric data that we have for our surveys, these two regimes overlap -- there is a range of parallaxes for which photometry dominates the goodness-of-fit $L_{sed}$ and fractional parallax uncertainty is small enough, so that $\pi'$ can be replaced by $\pi_0$ in $H_{0, 0}$.
To further understand these intermediate cases, we provide a more elaborate explanation of the properties of $L_{sed}$ and $H_{0,0}$ in \autoref{app:validity}.

With the addition of parallax and extinction priors we further constrain the stellar models that match the observations. In some cases these priors are not consistent with spectrophotometric data.
This can result in an increase in $L_{sed}$ for models used to build the PDFs and from that in the broadening of PDFs of all parameters, including log(age) and distance modulus.
A consequence of this is the increase in the uncertainties in the derived parameters.
This also decreases the best-model probability $p_{best}$ introduced in Paper I. Low values of $p_{best}$ generally indicate disagreement between constraints based on spectroscopy, photometry, parallax and extinction.

\section{Applications of the modified UniDAM}\label{sec:data}
The goals of this study are first, to provide an indication of the precision with which age and distance can be determined using Gaia end-of-mission (EoM) parallaxes and uncertainties and second, to re-derive distances and ages of stars for which TGAS parallaxes are available. To achieve the latter goal we cross-matched all spectroscopic surveys used in Paper I with TGAS, and run the updated UniDAM including parallax and extinction priors on the stars for which TGAS parallaxes are available.
\autoref{tbl:overlap} shows the number of TGAS stars contained in different surveys (see Paper I for a discussion of survey properties and quality cuts). The middle panels in Figures \ref{fig:uncertainties} and \ref{fig:unc2} show the number of stars in the complete survey and in the TGAS overlap as a function of measured distance modulus. For most surveys only a small fraction of all stars has TGAS counterpart and these stars are closer on average. The exception is GCS, for which some nearby stars were not included in TGAS, because they were above the brightness limit of TGAS. As in Paper I, we include GCS into our sample, although it is a photometric survey. In GCS narrow-band photometry is used to derive stellar parameters (\threeparams) with precisions comparable to those obtained with low-resolution spectroscopy.

In addition to the data used in Paper I, we added recently released LAMOST DR3\footnote{\url{dr3.lamost.org}} and TESS-HERMES DR1 \citep{TESS_HERMES}. For the APOGEE survey, we switched to using DR14 \citep{2017arXiv170709322A}. As before, 2MASS and AllWISE photometry was used. 
Small overlap with TGAS for most surveys is caused by the fact that TGAS contains primarily bright stars. There is no overlap between SEGUE \citep{SEGUE} and TGAS and we estimate  only what can be achieved with Gaia end-of-mission parallaxes for this survey. About one quarter of SEGUE stars have no 2MASS or AllWISE counterpart, and thus there is no photometry that can be used in UniDAM\footnote{We motivate the use of only infra-red photometry data in Paper I.}. These sources were excluded from analysis.
With deep spectroscopic surveys like SEGUE, LAMOST GAC and Gaia-ESO we will have to wait for later Gaia data releases to obtain parallaxes for the majority of stars.

We chose a prior on extinction as defined in \autoref{eq:extprior}. For this prior we took the mean value $A_0$ from Schlegel map \citep{1998ApJ...500..525S}. This map has a resolution of approximately $0.1$ degree. The variance of the extinction $\sigma_{A_0}^2$ for a given map cell was calculated as a variance of extinction values within one degree from the centre of that cell.

The prime goal is to show what to expect in terms of age and distance modulus precisions when using parallax priors with Gaia EoM precisions. For this, we simulated Gaia EoM data by taking parallax values from the TGAS or UniDAM catalogue and assigning Gaia EoM parallax uncertainties to these parallaxes. The distributions of uncertainties for distance and log(age) values derived using these simulated data are representative for what we expect to obtain with Gaia EoM data. Henceforth, we can estimate precisions with which ages and distances can be determined.

\begin{table*}[t]
    \centering
    \begin{tabular}{lrrc}
        \toprule
        Survey                                & \MyHead{2cm}{Total number of sources} & TGAS overlap & Reference  \\ \midrule
        APOGEE DR14                          & 157,322                 & 14,584       & \prefcount \\
        Gaia-ESO                              & 6,376                   & 67           & \prefcount \\
        GALAH DR1                            & 10,680                  & 7,919        & \prefcount \\
        GCS                                   & 13,565                  & 12,011       & \prefcount \\
        LAMOST DR3                           & 3,036,870               & 150,651      & \prefcount \\
        LAMOST GAC DR2 (Main sample)*        & 366,173                 & 541          & \prefcount \\
        LAMOST GAC DR2 (Very bright sample)* & 1,063,950               & 88,769       & \therefcount \\
        LAMOST-Cannon*                        & 444,784                 & 27,892       & \prefcount \\
        RAVE DR5                             & 440,913                 & 211,172      & \prefcount \\
        RAVE-on*                              & 491,349                 & 195,480      & \prefcount \\
        SEGUE**                              & 206,536                 & 0 & \prefcount \\
        TESS-HERMES DR1                      & 15,872                  & 5,928        & \prefcount \\ \midrule
        Total                                 &   3,888,134  & 402,732      &            \\ \bottomrule
    \end{tabular}
    \caption{Total number of sources and TGAS overlap for different surveys. *- LAMOST GAC, LAMOST-Cannon and RAVE-on were processed but not included into total, as they contain the same stars as LAMOST DR3 and RAVE DR5. ** - for SEGUE, we list the number of stars that have spectrophotometric parameters and Gaia DR1 counterpart.}
    \tablebib{\setcounter{refcount}{0}
        \pprefcount \cite{2017arXiv170709322A}; 
        \pprefcount \cite{GAIA_ESO};
        \pprefcount \cite{GALAH}; 
        \pprefcount \cite{GCS}; 
        \pprefcount \cite{LAMOST}; 
        \pprefcount \cite{2017arXiv170105409X}; 
        \pprefcount \cite{LAMOST_Cannon};    
        \pprefcount \cite{RAVE_DR5}; 
        \pprefcount \cite{RAVE_ON}; 
        \pprefcount \cite{SEGUE};
        \pprefcount \cite{TESS_HERMES}; 
    }\label{tbl:overlap}
\end{table*}

Overall, for each of the spectroscopic surveys we compiled five datasets as follows:
\begin{enumerate}
    \item Complete data of spectroscopic survey without parallaxes -- these are the same data as presented in Paper I.
    \item Subset of dataset 1, containing only sources that have a TGAS counterpart. For this dataset no parallax information was used. 
    \item Same as dataset 2, now using parallax data from TGAS\@.
    \item Same as dataset 3, now with parallax uncertainties as they are expected to be at the end of the Gaia mission according to the Gaia science performance guide\footnote{\url{http://www.cosmos.esa.int/web/gaia/science-performance}}, namely:
    \begin{equation} 
    \sigma_\pi = \left\{ 
    \begin{array}{l}
    10^{-5}\,\textrm{arcsec, if Gaia G-magnitude }\,m_G < 12^m; \\
    10^{-5 + (-12 + m_G) / 5.5}\,\textrm{arcsec, otherwise.} 
    \end{array}\right. \label{eq:eom_uncertainty}
    \end{equation}
    Here, $m_G$ is visible magnitude in Gaia optical $G$-band.
    \item Same as dataset 1, extended with parallaxes already obtained by UniDAM and assuming parallax uncertainties as they are expected to be at the end of the Gaia mission (see \autoref{eq:eom_uncertainty}). For stars in the spectroscopic survey the overlap with Gaia DR1 is used to extract values of $m_G$. These values are substituted into \autoref{eq:eom_uncertainty} to calculate Gaia EoM parallax uncertainty. There is a small (typically less than 3 percent) fraction of stars in each spectroscopic survey that does not have a counterpart in Gaia DR1 and these stars are not included into this dataset. We consider this difference in the sample of stars in dataset 1 and this dataset negligible for our purposes. Note that as the number of stars released in Gaia DR2 is expected to grow from 1.1 to 1.5 billion \citep{2017arXiv171010816K}, we expect that the fraction of stars in spectroscopic surveys without Gaia counterpart will further decrease.
\end{enumerate}

Summarizing, datasets 1 and 2 represent data without prior parallax information, dataset 3 represents data with the current state-of-the-art TGAS parallaxes and datasets 4 and 5 simulate Gaia EoM parallax precision. The last two datasets aim at demonstrating the improvements in the precision of age and distance that can be expected from including priors from parallaxes with Gaia EoM precision. 

Among datasets that include parallax data with Gaia EoM precision, dataset 4 has the advantage of having precise TGAS parallaxes, while dataset 5 contains a larger number of stars covering a large range in parallaxes, which better represents the content of Gaia end of mission data. As compared to the TGAS sample, this dataset contains more faint stars, for which Gaia EoM parallax uncertainty will be higher. Consequently, we expect distance and age uncertainties for all stars in a certain distance bin to be higher for datasets 5 than for dataset 4. In the last paragraph of \autoref{sec:new_method} we discussed that the addition of Gaia parallax and extinction values can lead to higher uncertainties in derived distance and age, in case the constraints provided by parallax and extinction priors and the spectrophotometric constraints are not consistent. Parallaxes used in dataset 5, however, are taken from UniDAM output for dataset 1, and not from Gaia. Therefore, the uncertainties derived for dataset 5 are in the limit of no mismatch between parallax constraints. Still, there can be a disagreement between extinction prior and constraints on extinction from spectrophotometric data, which can lead to higher uncertainties.

Following \autoref{eq:eom_uncertainty}, fainter stars will have larger parallax uncertainties and therefore larger uncertainties in the derived distances. Age uncertainty is also sensitive to stellar brightness, however in a less direct way: fainter stars typically have larger uncertainties in visible magnitudes and lower signal-to-noise ratio of observed spectra, and therefore larger uncertainties in spectroscopic parameters. Additionally, age uncertainty depends much more on the location of a star in the \HRD, due to the different rate of changes in $\teff$ and $\log g$ during different stages of stellar evolution. For example, during the main sequence evolutionary stage, $\teff$ and $\log g$ of a star are nearly constant, while around the turn-off point $\teff$ changes rapidly and on the red-giant branch $\log g$ changes rapidly. Therefore, the median measured age precisions for each dataset depend directly on the distribution of stars in that dataset. 
This implies that only datasets containing the same stars can be compared directly. Therefore, dataset 1, containing all survey stars can only be directly compared to dataset 5 (here we neglect the small difference in the number of stars between datasets 1 and 5); and dataset 2, containing only the TGAS overlap can only be directly compared to datasets 3 and 4.

\section{Results and discussion}
We apply the updated UniDAM, as described in the \autoref{sec:method}, to all datasets for each spectroscopic survey presented in \autoref{sec:data}. We publish the results for dataset 1 for new surveys (TESS-HERMES \citep{TESS_HERMES}, LAMOST DR3 \mbox{\citep{LAMOST}} and APOGEE DR14 \citep{2017arXiv170709322A}), thus extending the number of stars in our catalogue as compared to the catalogue in the Paper 1 to nearly 4 million stars. We also publish results for dataset 3 for all surveys along with this paper. These results contain improved log(age) and distance estimates for about 400,000 stars.

In this section we compare the precision of distance modulus and log(age) for different datasets. The quantitative behaviour of the relation of distance modulus uncertainty to distance modulus in all surveys depends on the content of each dataset for each spectroscopic surveys, hence we list and discuss our results for every dataset for each survey. 
We first consider distance modulus precision in \autoref{sec:mud_precision} and then proceed to log(age) precision in \autoref{sec:age_precision}.

\subsection{Distance modulus precision}\label{sec:mud_precision}
Distance modulus precision is closely related to parallax precision:
for a value of parallax uncertainty $\sigma_{\pi_0}$, the corresponding distance modulus uncertainty $\sigma_{\mu_d}$ is a function of distance modulus $\mu_d$ and can be expressed as:
\begin{equation}
\sigma_{\mu_d} = \frac{5\sigma_{\pi_0} 10^{1 + 0.2 \mu_d}}{\ln 10}.\label{eq:propagate}
\end{equation}
The above equation is valid only approximately, as the distance modulus uncertainties become asymmetric, when propagated from the symmetric parallax uncertainties \citep[see][]{1998A&A...340L..35K, 2016ApJ...832..137A}.
From \autoref{eq:propagate} it follows that the distance modulus uncertainty depends on two values: first, the parallax uncertainty, and second, the value of the distance modulus itself. For Gaia EoM, parallax uncertainty estimates depend on visible magnitudes for stars fainter than $12^m$ in G-band (see \autoref{eq:eom_uncertainty}).
The distance modulus uncertainty derived from spectrophotometric data is much less sensitive to visible magnitudes.
Once a star is bright enough to be accessible to spectrophotometric observations, uncertainties in the measured parameters, and thus the derived distance uncertainty, will not depend directly on the distance. Therefore, we can expect that for distant stars (with $\mu_d \gtrsim 15^m$) the spectrophotometric data constrain the distance modulus more than parallax data. Thus, the benefits of using the parallax priors for surveys containing primarily nearby stars (like GCS and TESS-HERMES) will be higher than those for deep surveys containing distant stars, which are on average fainter. For deep surveys, like APOGEE, SEGUE, LAMOST and future surveys like 4MOST \citep{4MOST} and WEAVE \citep{WEAVE}, there will be a high percentage of stars for which spectrophotometric information will improve distance modulus uncertainties.

\begin{figure}
    \myimagesmall{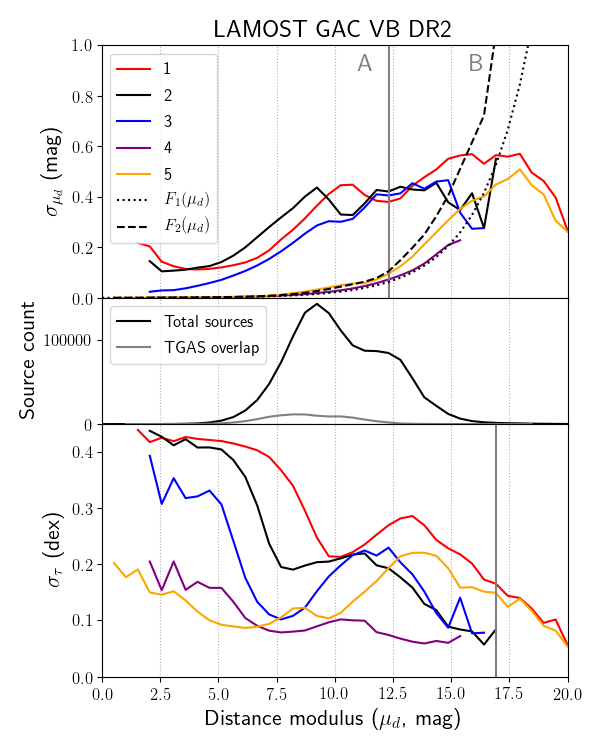}
    \caption{Distance modulus uncertainties $\sigma_{\mu_d}$ (top panel) and log(age) uncertainties $\sigma_{\tau}$ (bottom panel) as functions of the distance modulus. Lines show median values in $0.^m5$ distance modulus bins. Colours are for datasets: dataset 1 -- complete survey (red); dataset 2 -- TGAS overlap (black); dataset 3 -- TGAS overlap with TGAS parallaxes (blue); dataset 4 -- TGAS parallaxes with EoM precisions (magenta); dataset 5 -- UniDAM parallax with EoM precisions (orange). We show only distance modulus bins with at least five stars, to reduce the noise. Therefore, the distance modulus range presented can differ for different datasets, especially for surveys with small overlap with TGAS\@.
        The dotted black line shows $F_1(\mu_d)$, the dashed black line shows $F_2(\mu_d)$ (as described in \autoref{sec:mud_precision}). 
        Vertical grey lines and grey labels at the top panel mark borders between ranges A, B and C (see \autoref{sec:mud_precision} for details). Vertical grey line at the bottom plot marks maximum distance modulus for which the use of parallaxes gives at least $10\%$ improvement in log(age) uncertainty (see \autoref{sec:age_precision}).
        The middle panel shows the total number of stars in the survey (black) and the number of stars in overlap with TGAS (grey) as a function of distance modulus.}
    \label{fig:uncertainties}
\end{figure}

The top panels of Figs.~\ref{fig:uncertainties} and~\ref{fig:unc2} show median distance modulus uncertainties $\sigma_{\mu_d}$ as functions of measured distance modulus $\mu_d$.
The results for dataset 3 show that the use of TGAS parallaxes provides a substantial improvement in the precision of distance modulus estimation as compared to dataset 2, which contains the same data without parallaxes, for distance moduli up to $\mu_d = 10^m$. 

For illustrative purpose and to better understand what limits the distance modulus uncertainty we produced two approximations for $\sigma_{\mu_d}(\mu_d)$ function. They are shown in each top panel of \autoref{fig:uncertainties} and \autoref{fig:unc2} with black dashed and dotted lines. 
The first approximation $F_1(\mu_d)$ is obtained using \autoref{eq:propagate} to propagate $10^{-5}$\,arcsec parallax uncertainties expected for stars brighter than $m_G = 12^m$ (see \autoref{eq:eom_uncertainty}) to distance modulus uncertainties. This provides an approximation of the resulting precision when TGAS parallaxes are used with Gaia EoM parallax precisions (dataset 4), because almost 90 percent of TGAS stars are brighter than $m_G = 12^m$. This approximation is indeed representative of the measured precision, except for the most distant ($\mu_d > 15^m$) bins in the RAVE surveys: at these distances measured parallax values become close to typical parallax uncertainty for RAVE stars ($10^{-5}$ arcsec) and spectrophotometric constraints improve distance estimations.

When parallax values are taken from UniDAM and parallax uncertainties are taken from Gaia EoM uncertainty prediction (dataset 5) the above approximation is not working, because for stars in that dataset that are fainter than $m_G = 12^m$ the parallax uncertainty will be larger than $10^{-5}$\,arcsec. To produce an approximation for these cases we took for each star its $\mu_d$ as derived by UniDAM and additionally calculated $\sigma_{\pi_0}$ using \autoref{eq:eom_uncertainty} and visible magnitude $m_G$ from Gaia DR1. Values of  $\sigma_{\pi_0}$ were than propagated to distance modulus uncertainty using \autoref{eq:propagate}. We then obtained the mean distance modulus uncertainty as a function of distance modulus.  The resulting function $F_2(\mu_d)$ is shown in Figs. \ref{fig:uncertainties} and \ref{fig:unc2} with black dashed lines.

To illustrate how results for dataset 5 compare with parallax-only approximation $F_2(\mu_d)$ and results for spectrophotometry-only dataset 1, we can define three distance ranges of interest: 
\begin{enumerate}[label=Range \Alph*, itemindent=*]
    \item -- ``parallax dominated'', where distance modulus is almost entirely defined by parallax. We define it as a range where the improvement in the distance modulus uncertainty for dataset 5 is less than $10\%$ as compared to our approximation $F_2$  .
    \item -- ``intermediate'', where distance modulus is improved by the use of parallax. We define it as a range that spans from the upper limit of the range A to the point where the improvement in the distance modulus uncertainty for dataset 5 is less than $10\%$ as compared to the dataset 1 (in which only spectrophotometric data is used).
    \item -- ``spectrophotometry dominated'', where parallax has almost no influence on distance modulus uncertainty. We define it as a range where the improvement in distance modulus uncertainty for dataset 5 is less than $10\%$ as compared to the dataset 1 (in which only spectrophotometric data is used).
\end{enumerate}
These ranges are labelled in the top panels of \autoref{fig:uncertainties} and \autoref{fig:unc2} with large letters A, B and C. Range borders are marked with vertical grey lines.

In \autoref{tbl:improvement} we show fractions of stars in each survey that fall into ranges A, B and C, as well as locations of range borders. These fractions depend on the distribution of stars in both visible and absolute magnitudes and thus are very different from survey to survey. For GALAH, GCS, RAVE, RAVE-on and TESS-HERMES the majority (over $95\%$) of stars falls into the range A, where Gaia parallaxes define the distance modulus. 
For other surveys spectrophotometric constraints play a role for a large portion of stars: for LAMOST-based surveys, Gaia-ESO and APOGEE 20 to 80 percent stars are in the range B, which means that spectrophotometry constraints give at least 10 percent improvement over pure-Gaia distances. Range B extends from distance modulus of approximately $11^m$ to $16^m$, or from 1.5 to 15 kiloparsecs. 
A large portion of stars is in range C for Gaia-ESO ($6\%$), APOGEE ($7\%$) and SEGUE ($37\%$). This means that Gaia parallax will give almost no improvement in distance modulus for these stars. Also, for SEGUE only a few stars are in range A. This is because SEGUE stars are faint, and Gaia parallax uncertainties will be higher for them than for stars in other surveys. We expect for future surveys to contain a large portion of stars in ranges B and C, and thus we will need a combination of spectrophotometric data and Gaia parallaxes to derive best possible distance estimates for them.

\begin{table*}
    \begin{center}
        \begin{tabular}{lrrrrr}
            \toprule
            Survey            & \MyHead{1.8cm}{Range A \\ (\% of stars)} & \MyHead{1.8cm}{Range A-B \\ border} & \MyHead{1.8cm}{Range B \\ (\% of stars)} & \MyHead{1.8cm}{Range B-C \\ border} & \MyHead{1.8cm}{Range C \\ (\% of stars)} \\ \midrule
            APOGEE DR14 &              15.3 &         $11^m.3$ &              77.9 &         $15^m.4$ &               6.8 \\
            GALAH DR1 &              97.6 &         $13^m.8$ &               2.4 &         $\infty$ &               -- \\
            GCS &              99.9 &         $11^m.3$ &               0.1 &         $\infty$ &               -- \\
            Gaia-ESO DR2 &              28.2 &         $10^m.8$ &              65.9 &         $15^m.9$ &               5.8 \\
            LAMOST DR3 &              62.9 &         $11^m.3$ &              34.1 &         $15^m.9$ &               3.0 \\
            LAMOST GAC DR2 &              63.3 &         $12^m.3$ &              35.9 &         $16^m.9$ &               0.7 \\
            LAMOST GAC VB DR2 &              78.3 &         $12^m.3$ &              21.7 &         $\infty$ &               -- \\
            LAMOST-Cannon &              51.7 &         $12^m.3$ &              44.8 &         $15^m.4$ &               3.5 \\
            RAVE DR5 &              97.9 &         $15^m.4$ &               2.0 &         $19^m.0$ &               -- \\
            RAVE-on &              95.9 &         $14^m.4$ &               4.1 &         $17^m.4$ &               0.1 \\
            SEGUE &               2.8 &          $9^m.2$ &              60.5 &         $13^m.8$ &              36.8 \\
            TESS-HERMES DR1 &              99.8 &         $11^m.3$ &               0.2 &         $\infty$ &               -- \\
            \midrule
            Total      & 61.7 & -- & 33.7 & -- &  4.6 \\
            \bottomrule
        \end{tabular}
        \caption{Ranges of distance modulus $\mu_d$ improvement from the use of Gaia end of mission parallax priors. First column shows survey label (see \autoref{tbl:overlap} for references); Second, fourth and sixth columns show percentage of survey stars within ranges A, B and C (see \autoref{sec:mud_precision} for definitions of these ranges). Third and fifth column show distance modulus values at range borders. Last row shows what fraction of the total number of stars in our sample (see the last row of \autoref{tbl:overlap}) falls within each range.  *- LAMOST GAC, LAMOST-Cannon and RAVE-on were processed but not included into total, as they contain the same stars as LAMOST DR3 and RAVE DR5.}
        \label{tbl:improvement}
    \end{center}
\end{table*}

In \autoref{tbl:results} we give another view on results by listing the median uncertainty per-dataset for each survey. 
The first section of \autoref{tbl:results} shows median distance modulus uncertainties for each dataset for each spectroscopic survey. 
Datasets 1 and 2 are built using only spectrophotometric data, and the small difference between median uncertainty values for these datasets is caused by the fact that stars in TGAS overlap are typically brighter. For these stars input spectrophotometric parameters are generally more precise than those for fainter survey stars at the same distance, and hence distance modulus and log(age) uncertainties are smaller.

\begin{table*}
    \begin{center}
        \begin{tabular}{lrrrrr}
            \toprule
            Survey              & \MyHead{1.8cm}{Complete survey (dataset 1)} & \MyHead{1.8cm}{TGAS overlap (dataset 2)} & \MyHead{1.8cm}{TGAS parallaxes (dataset 3)} & \MyHead{1.8cm}{TGAS parallaxes, EoM precision (dataset 4)} & \MyHead{1.8cm}{UniDAM parallaxes, EoM precision (dataset 5)} \\ \midrule
            Parallaxes:         &                                           - &                                        - &                                        TGAS &                                                       TGAS &                                                       UniDAM \\
            Parallax precision: &                                           - &                                        - &                                        TGAS &                                                        EoM &                                                          EoM \\
            Survey data:        &                                        Full &                                                          \multicolumn{3}{c}{TGAS overlap}                                                           &                                                        Full* \\ \midrule
            \multicolumn{6}{c}{1. Median distance modulus uncertainty ($\sigma_{\mu_d}$) (mag) }                                                                                                  \\ \midrule
            APOGEE DR14         &                                       0.233 &                                    0.207 &                                       0.177 &                                                      0.023 &                                                        0.094 \\
            GALAH DR1           &                                       0.547 &                                    0.549 &                                       0.283 &                                                      0.014 &                                                        0.012 \\
            GCS                 &                                       0.287 &                                    0.288 &                                       0.046 &                                                      0.002 &                                                        0.002 \\
            Gaia-ESO DR2        &                                       0.232 &                                    0.229 &                                       0.108 &                                                      0.005 &                                                        0.106 \\
            LAMOST DR3          &                                       0.287 &                                    0.123 &                                       0.100 &                                                      0.011 &                                                        0.048 \\
            LAMOST GAC DR2      &                                       0.375 &                                    0.260 &                                       0.166 &                                                      0.011 &                                                        0.119 \\
            LAMOST GAC VB DR2   &                                       0.327 &                                    0.313 &                                       0.209 &                                                      0.011 &                                                        0.029 \\
            LAMOST-Cannon      &                                       0.352 &                                    0.327 &                                       0.248 &                                                      0.023 &                                                        0.076 \\
            RAVE DR5            &                                       0.377 &                                    0.372 &                                       0.226 &                                                      0.012 &                                                        0.016 \\
            RAVE-on             &                                       0.373 &                                    0.376 &                                       0.219 &                                                      0.013 &                                                        0.019 \\
            SEGUE               &                                       0.497 &                                      -  &                                         - &                                                        - &                                                        0.231 \\
            TESS-HERMES DR1     &                                       0.397 &                                    0.358 &                                       0.217 &                                                      0.009 &                                                        0.011 \\ \midrule
            \multicolumn{6}{c}{2. Median log(age) uncertainty ($\sigma_{\tau}$) (dex)}                                                                                                       \\ \midrule
            APOGEE DR14       &           0.206 &        0.188 &           0.169 &                                0.088 &                                  0.138 \\
            GALAH DR1         &           0.289 &        0.289 &           0.157 &                                0.088 &                                  0.116 \\
            GCS               &           0.134 &        0.133 &           0.073 &                                0.070 &                                  0.083 \\
            Gaia-ESO DR2      &           0.237 &        0.223 &           0.095 &                                0.073 &                                  0.167 \\
            LAMOST DR3        &           0.302 &        0.089 &           0.074 &                                0.063 &                                  0.172 \\
            LAMOST GAC DR2    &           0.312 &        0.242 &           0.153 &                                0.083 &                                  0.213 \\
            LAMOST GAC VB DR2 &           0.275 &        0.224 &           0.146 &                                0.092 &                                  0.122 \\
            LAMOST-Cannon     &           0.224 &        0.224 &           0.198 &                                0.116 &                                  0.158 \\
            RAVE DR5          &           0.234 &        0.215 &           0.158 &                                0.092 &                                  0.120 \\
            RAVE-on           &           0.245 &        0.227 &           0.167 &                                0.093 &                                  0.133 \\
            SEGUE             &           0.349 &          - &             - &                                  - &                                  0.303 \\
            TESS-HERMES DR1   &           0.242 &        0.185 &           0.093 &                                0.054 &                                  0.092 \\ \bottomrule
        \end{tabular}
        \caption{Median uncertainties for distance modulus ($\sigma_{\mu_d}$) and log(age) ($\sigma_{\tau}$) for five datasets derived for each survey. First three rows describe data used for each dataset: which values of parallaxes and parallax uncertainties were used and what part of the survey was used (see for more details \autoref{sec:data}). * - there is a small (less than $3\%$) fraction of stars in each survey that do not have Gaia DR1 counterpart and are therefore not included into dataset 5 }
        \label{tbl:results}
    \end{center}
\end{table*}

The use of TGAS parallaxes for dataset 3 increases the precision in the distance modulus by about one third. The exception here is GCS, for which TGAS parallaxes increase distance modulus precision almost by a factor of five. This is caused by the fact that GCS contains nearby \textit{Hipparcos} stars, for which parallaxes in TGAS have high-precision.

For datasets 4 and 5, which are constructed using expected Gaia EoM parallax precisions, the median distance modulus uncertainties are on the order of $0^m.1$ or even $0^m.01$. The actual value depends primarily on the distance distribution of stars in the spectroscopic survey which means that for deeper surveys parallax uncertainties will be larger, leading to larger  distance modulus uncertainties. Similarly, dataset 5 as compared to dataset 4, contains more distant stars that are fainter on the average. Therefore, median  distance modulus uncertainties  for dataset 5 are in most cases larger than those for dataset 4.

\subsection{log(age) precision}\label{sec:age_precision}
The effect of the parallax priors in UniDAM on the log(age) uncertainty is presented in the lower panels of Figures~\ref{fig:uncertainties} and~\ref{fig:unc2} and in Tables~\ref{tbl:ageimprovement} and~\ref{tbl:results}. This effect is not straightforward to quantify. Parallax data constrain the absolute magnitude of the star, which can have impact on the log(age) estimate. This improvement varies along the \HRD, with improvements being larger for lower main-sequence stars, and smaller or close to zero for the main sequence turn-off region. 
In the turn-off region age depends less on absolute magnitudes and more on temperature, therefore, little or no further improvement of log(age) estimates can be made by adding parallax data.

The decrease in log(age) uncertainties with distance modulus that is present in some surveys is caused by the fact that log(age) uncertainties are much higher for lower main-sequence stars than for turn-off stars and giants. At the same time, at a given distance, main-sequence stars are harder to detect than giants, because they are intrinsically fainter. This causes the fraction of observed main-sequence stars per distance modulus bin to decrease with distance modulus. Hence, the median log(age) uncertainty per bin decreases with distance modulus too. For APOGEE and LAMOST-Cannon surveys, which contain preferentially giants, this decreasing trend is absent.

With \autoref{tbl:ageimprovement} we illustrate where we can improve log(age) uncertainties by using TGAS and Gaia EoM parallaxes. We do that by calculating the maximum distance at which the parallax prior has almost no influence on log(age) uncertainty. We define this for the TGAS sample as a range where the uncertainty for datasets 3 is smaller than $90\%$ of the uncertainty for dataset 2 (in which only spectrophotometric data are used). Similarly, for Gaia EoM parallaxes, we compare datasets 5 and 1. In both cases, we provide the fractions of stars that are within the listed ranges. In the bottom panels of \autoref{fig:uncertainties} and \autoref{fig:unc2} we show with a vertical grey line the maximum distance modulus for which log(age) uncertainty for dataset 5 is smaller than $90\%$ of the uncertainty for dataset 1 (see fourth column of \autoref{tbl:ageimprovement}).
For GCS the improvement in log(age) is more than $10\%$ for all stars, and this survey is thus not listed in the \autoref{tbl:ageimprovement}. 
We observe log(age) estimate improvements from the use of TGAS parallaxes for stars with distance moduli up to $\approx 11^m$. Depending on the survey there are one half to over three quarters of stars in this range. The exceptions are APOGEE and LAMOST-Cannon surveys, for which the improvement is seen for less than $40\%$ of stars -- stars in the overlap with TGAS for these surveys are more distant on average, and thus fractional parallax uncertainties are higher for them, which causes less improvements in log(age).

When we consider the effect of Gaia EoM parallax priors, two groups of surveys can be seen. For surveys focussing on brighter stars, like GALAH, GCS, LAMOST GAC VB, RAVE surveys and TESS-HERMES, log(age) estimates will improve for almost all stars. The maximum distance modulus $\mu_d$ listed in \autoref{tbl:ageimprovement} for these surveys is therefore not very informative, as it reflects the distance to most distant stars in the survey. For the group of surveys containing fainter stars, to which APOGEE, Gaia-ESO and LAMOST surveys (excluding LAMOST GAC VB) belong, the fraction of stars is lower and ranges from $80$ to $95\%$. The maximum distance modulus listed in \autoref{tbl:ageimprovement} for these surveys is ranging from $13^m.85$ to $14^m.87$, or between 6 and 9.5 kiloparsecs. Similar values can also be assumed for future surveys.

The low number of stars with improvement in log(age) uncertainty for SEGUE is explained by the fact that the majority of stars in this survey are faint and distant. Hence, parallax uncertainties for such stars will be higher than for stars in other surveys, which leads to less improvement is log(age) estimates.

The second part of \autoref{tbl:results} lists median values of the log(age) uncertainty for each dataset for each survey.
The typical median log(age) uncertainty for dataset 3, in which TGAS parallaxes are included, is $0.17$\,dex. That corresponds to a fractional uncertainty of about $35\%$ in age, compared to $0.22$\,dex log(age) uncertainty (or over $50\%$ age uncertainty) calculated without parallax for datasets 1 and 2. 

When parallaxes with Gaia EoM uncertainties are used (for datasets 4 and 5), median log(age) uncertainties further decrease -- to a typical value of $0.1$\,dex (or about $25\%$ in age). Median log(age) uncertainties are not lower than $0.063$\,dex (which corresponds to $15\%$ uncertainty in age) for dataset 4 and $0.083$\,dex ($18\%$ uncertainty in age) for dataset 5. At these precisions we are limited by the uncertainties in the spectroscopic, photometric and extinction measurements and not by the parallax precision. 

Overall, we deduce that Gaia EoM parallax priors will allow to improve log(age) uncertainties for at least $80\%$ stars in existing surveys. This value is less than the fraction of stars, for which an improvement in distance modulus uncertainty is predicted. This reflects facts that the parallax prior directly constrains distance modulus, while it constrains log(age) only through absolute magnitudes. The typical log(age) uncertainty values is expected to be around $0.1$\,dex.

\begin{table*}
    \begin{center}
\begin{tabular}{lrrrr}
	\toprule
	\multirow{2}{*}{Survey} &    \multicolumn{2}{c}{TGAS (dataset 2 / dataset 3)}     &  \multicolumn{2}{c}{Gaia EoM (dataset 1 / dataset 5)}   \\
	                        & Maximum $\mu_d$ & Fraction (\%) & Maximum $\mu_d$ & Fraction (\%) \\ \midrule
     APOGEE DR14 &           $9^m.7$ &              25.7 &          $14^m.4$ &               80.4 \\
    GALAH DR1 &          $11^m.8$ &              85.1 &          $\infty$ &              100.0 \\
    Gaia-ESO DR2 &          $10^m.3$ &              84.4 &          $14^m.4$ &               82.7 \\
    LAMOST DR3 &           $8^m.7$ &              53.8 &          $14^m.9$ &               94.4 \\
    LAMOST GAC DR2 &          $10^m.8$ &              76.8 &          $13^m.8$ &               85.3 \\
    LAMOST GAC VB DR2 &          $10^m.3$ &              71.8 &          $16^m.9$ &               99.7 \\
    LAMOST-Cannon &          $10^m.3$ &              38.5 &          $14^m.9$ &               94.3 \\
    RAVE DR5 &          $11^m.3$ &              78.9 &          $\infty$ &              100.0 \\
    RAVE-on &          $10^m.8$ &              74.3 &          $17^m.4$ &               99.9 \\
    SEGUE &                 - &                 - &          $10^m.8$ &               19.3 \\
    TESS-HERMES DR1 &          $10^m.3$ &              97.9 &          $\infty$ &              100.0 \\ \bottomrule
\end{tabular}
\caption{Maximum distance modulus $\mu_d$ and a fraction of stars, for which the use of parallaxes and spectrophotometric data gives at least $10\%$ improvement in log(age) uncertainties as compared to those obtained with spectrophotometric data alone. Second and third columns are the result of comparing datasets 2 and 3 (without and with TGAS parallaxes used). Fourth and fifth columns are the result of comparing datasets 1 and 5 (without and with Gaia EoM parallaxes used).}
   \label{tbl:ageimprovement}
\end{center}
\end{table*}

\section{Summary and conclusions}
In this work we show that using the combination of Gaia end-of-mission parallax data and spectrophotometric data in the isochrone fitting we will reduce the distance modulus uncertainties to $0^m.1$ or even $0^m.01$ while log(age) uncertainties will decrease to about $0.1$\,dex. To this end we included Gaia parallax data into isochrone fitting in a consistent way. UniDAM \citep{Paper1} is updated to incorporate Gaia parallax measurements and Schlegel extinction data. With this updated tool, we calculate values of log(age) and distance for stars in public spectroscopic surveys that have a TGAS counterpart. The new catalogue contains distance and age estimates for over $400\,000$ stars, distributed over a large portion of the sky. The improvements are most substantial for distance modulus, which is directly related to parallax -- for the majority of stars in the TGAS overlap distance modulus precision is dominated by parallax precision.  For log(age) the typical uncertainty decreases by one third from $0.22$\,dex without parallaxes to $0.17$\,dex with parallaxes. 

When parallax priors with Gaia end of mission quality will be used in the isochrone fitting,
distance modulus uncertainty will be limited by parallax precision for distance moduli up to $10^m - 12^m$, or distances of 1 to 3 kiloparsecs, depending on the survey content. Beyond this range, Gaia parallaxes will still improve our distance modulus estimates -- up to a distance modulus of at least $14^m.36$, or 7.5 kiloparsecs. The impact of the use of parallaxes will be smaller for deep surveys which contain fainter stars, like APOGEE, SEGUE and to some extent Gaia-ESO. In the worst case of SEGUE, for one third of stars we expect only marginal improvement if any in distance modulus. 

Upon including parallax priors, the median log(age) uncertainties will be typically around $0.1$\,dex -- more than a factor of two better than log(age) uncertainties obtained without parallax data. We find that median uncertainties reach a minimum value at $0.083$\,dex, which is caused by our uncertainties in spectroscopic parameters, photometry and extinction values. We expect improvements in log(age) uncertainties of at least $10\%$ for stars with distance moduli up to approximately $14^m$ -- with a majority of stars (at least $80\%$) falling into this range. The only exception here is SEGUE: age is poorly constrained by isochrone fitting for faint main sequence stars of this catalogue, and  parallax information does not help to overcome this difficulty, because parallax uncertainties for these stars are also expected to be larger than for stars in other surveys.

It is important to determine log(age) and distance modulus in a consistent manner, even if the precision of the latter is dominated by parallax precision, because values of log(age) and distance modulus are correlated: an offset in distance modulus between a pure spectrophotometric estimate and the one that uses a parallax prior will cause an offset in log(age) estimates.

We are ready to use Gaia DR2 data as soon as it will become public. Once Gaia DR2 parallaxes are available we will be able to further improve distance modulus and log(age) estimates for the majority of stars in spectroscopic surveys. 

\section*{Acknowledgements}
The research leading to the presented results has received funding from the European Research Council under the European Community's Seventh Framework Programme (FP7/2007- 2013)/ERC grant agreement (No 338251, StellarAges). 

This research has made use of the VizieR catalogue access tool, CDS, Strasbourg, France. This research made use of Astropy, a community-developed core Python package for Astronomy \citep{2013A&A...558A..33A} This research made use of matplotlib, a Python library for publication quality graphics \citep{Hunter:2007} This research made use of SciPy \citep{SciPy} This research made use of TOPCAT, an interactive graphical viewer and editor for tabular data \citep{2005ASPC..347...29T} This publication makes use of data products from the Two Micron All Sky Survey, which is a joint project of the University of Massachusetts and the Infrared Processing and Analysis Center/California Institute of Technology, funded by the National Aeronautics and Space Administration and the National Science Foundation. Funding for SDSS-III has been provided by the Alfred P. Sloan Foundation, the Participating Institutions, the National Science Foundation, and the U.S. Department of Energy Office of Science. The SDSS-III web site is http://www.sdss3.org/. SDSS-III is managed by the Astrophysical Research Consortium for the Participating Institutions of the SDSS-III Collaboration including the University of Arizona, the Brazilian Participation Group, Brookhaven National Laboratory, University of Cambridge, Carnegie Mellon University, University of Florida, the French Participation Group, the German Participation Group, Harvard University, the Instituto de Astrofisica de Canarias, the Michigan State/Notre Dame/JINA Participation Group, Johns Hopkins University, Lawrence Berkeley National Laboratory, Max Planck Institute for Astrophysics, Max Planck Institute for Extraterrestrial Physics, New Mexico State University, New York University, Ohio State University, Pennsylvania State University, University of Portsmouth, Princeton University, the Spanish Participation Group, University of Tokyo, University of Utah, Vanderbilt University, University of Virginia, University of Washington, and Yale University. This publication makes use of data products from the Wide-field Infrared Survey Explorer, which is a joint project of the University of California, Los Angeles, and the Jet Propulsion Laboratory/California Institute of Technology, and NEOWISE, which is a project of the Jet Propulsion Laboratory/California Institute of Technology. WISE and NEOWISE are funded by the National Aeronautics and Space Administration Guoshoujing Telescope (the Large Sky Area Multi-Object Fiber Spectroscopic Telescope LAMOST) is a National Major Scientific Project built by the Chinese Academy of Sciences. Funding for the project has been provided by the National Development and Reform Commission. LAMOST is operated and managed by the National Astronomical Observatories, Chinese Academy of Sciences. Funding for RAVE has been provided by: the Australian Astronomical Observatory; the Leibniz-Institut fuer Astrophysik Potsdam (AIP); the Australian National University; the Australian Research Council; the French National Research Agency; the German Research Foundation (SPP 1177 and SFB 881); the European Research Council (ERC-StG 240271 Galactica); the Istituto Nazionale di Astrofisica at Padova; The Johns Hopkins University; the National Science Foundation of the USA (AST-0908326); the W. M. Keck foundation; the Macquarie University; the Netherlands Research School for Astronomy; the Natural Sciences and Engineering Research Council of Canada; the Slovenian Research Agency; the Swiss National Science Foundation; the Science \& Technology Facilities Council of the UK; Opticon; Strasbourg Observatory; and the Universities of Groningen, Heidelberg and Sydney. The RAVE web site is at \url{https://www.rave-survey.org}. Based on data products from observations made with ESO Tele-scopes at the La Silla Paranal Observatory under programme ID 188.B-3002. These data products have been processed by the Cambridge Astronomy Survey Unit (CASU) at the Institute of Astronomy, University of Cambridge, and by the FLAMES/UVES reduction team at INAF/Osservatorio Astrofisico di Arcetri. These data have been obtained from the Gaia-ESO Survey Data Archive, prepared and hosted by the Wide Field Astronomy Unit, Institute for Astronomy, University of Edinburgh, which is funded by the UK Science and Technology Facilities Council.

\appendix
\section{Minimising new goodness-of-fit}\label{app:A}
In \autoref{sec:method} we introduced goodness-of-fit $L_{sed}$ (\autoref{eq:lsed}). As in Paper I, by minimisation $L_{sed}$ we find the optimal values of $\mu_d$ and $A_K$ for each model that fits the stellar spectral parameters. Minima can be found as the solution of a system of equations:
\begin{equation}
\large{
  \begin{cases}
    \deriv{L_{sed}}{\mu_d} &= 0 \\
    \deriv{L_{sed}}{A_K} &= 0
  \end{cases}}.
\end{equation}
Calculating $\deriv{L_{sed}}{\mu_d}$ and $\deriv{L_{sed}}{A_K}$ and setting them to zero we obtain the following system of equations:
\begin{equation}\label{eq:system}
\left\{
\begin{array}{rrl}
\sum_\lambda \frac{- \mags}{\sigma_m^2} - 0.2 \ln 10 \left(\frac{10^{-0.2 \mu_d - 1} (10^{-0.2 \mu_d - 1} - \pi_0) }{\sigmapi^2} + 2 \right) = 0 \\
\sum_\lambda \frac{-C_\lambda \mags}{\sigma_m^2} + B \cdot \frac{A_K - A_0}{\sigma_{A_0}^2} = 0 \\
\end{array}\right.,
\end{equation}
where $B = 1$ if $A_K > A_0$ and $B=0$ otherwise.
This system is non-linear in $\mu_d$ and has to be solved numerically.
As before, we are confident that system~\ref{eq:system} has only one solution, because its second equation is linear and because the non-linear first equation has a left side that increases monotonically with $\mu_d$.
The solution of this system gives us the optimal $\mu_d$ and $A_K$ for each model, which can be used to produce PDFs of distance modulus and extinction.

\section{Method validity analysis}\label{app:validity}
Each model contributes to the distance modulus PDF a component that is defined by \autoref{eq:lsed}. As it was shown in Paper I, without parallaxes these components have exactly the same Gaussian shape, which width is defined by the Hessian matrix. When parallaxes are included, this is no longer the case. While $\frac{\partial^2 L_{sed}}{\partial {A_K}^2}$ and $\frac{\partial^2 L_{sed}}{\partial A_K \partial \mu_d}$ remain the same, the expression for $\dLdmu$ becomes much more complex (see \autoref{eq:lsed_2nd_der}):
\begin{equation}
H_{0,0} = \dLdmu = (0.2\ln(10))^{2} \cdot \frac{\pi\left(2\pi - \pi_0 \right) }{\sigmapi^2}+\sum_\lambda \frac{1}{\sigma_{m_\lambda}^2} 
\label{eq:app_lsed_2nd_der}
\end{equation}

There are two difficulties arising from \autoref{eq:app_lsed_2nd_der}. Namely, the first summand can be negative, which decreases the value of $H_{0, 0}$ as compared to a case when no parallax is included. This means that an addition of the parallax information can increase the width of the Gaussian contribution of some models to PDFs, broadening the PDFs. The second difficulty is that \autoref{eq:app_lsed_2nd_der} and therefore, the width of the Gaussian contribution of some models to PDFs now depends in addition to $\sigma_{m_\lambda}$, on $\sigmapi$ and $\pi_0$, which are constants for a given star, as well as on the parallax value $\pi$ for a given model. We show that these difficulties can be solved by proving two following statements:
\begin{description}
    \item [Statement 1:] First summand ($S=(0.2\ln(10))^{2} \cdot \frac{\pi\cdot\left(2\pi - \pi_0 \right) }{\sigmapi^2}$) of $\dLdmu$ can be negative, but in this case its contribution will be small, and thus $\dLdmu$ is never significantly smaller than $\sum_\lambda \frac{1}{\sigma_{m_\lambda}^2}$. The latter sum is the exact value of $\dLdmu$ when no parallax information is included.
    \item [Statement 2:] $\pi$ can be replaced by $\pi_0$ in $S$, as $\pi$ is close to $\pi_0$ in all cases where the contribution of $S$ to $\dLdmu$ is substantial. 
\end{description}

\subsection{Proof of a Statement 1}
The first summand ($S$) is negative for $0 < \pi < \pi_0 / 2$, and reaches its minimum value at $\pi = \pi_0 / 4$. This minimum value is thus $(0.2\ln(10))^{2} \pi_0^2 / (8 \sigmapi^2)$.

At the same time, we consider only small values of $L_{sed}$, as otherwise the model can be considered to be unreliable. UniDAM produces a solution only if the chi-square probability for the best-fitting model is more than $3\%$.
We require the chi-square probability derived from $L_{sed}$ for the given model to be at least $0.1\%$ of that for the best model (otherwise its contribution to the PDF will be negligible). This implies that $L_{sed} \lesssim 25$ if the number of degrees of freedom is 7. The number of degrees of freedom in this case is the number of frequency bands for which visible magnitudes are available plus two -- for extinction and parallax.

If we require $L_{sed} < 25$, then for $\pi = \pi_0 / 2$ we get:
\begin{equation}
\frac{(\pi - \pi_0)^2}{2\sigmapi^2} = \frac{\pi_0^2}{8 \sigmapi^2} < L_{sed} < 25
\end{equation}
and thus if $\pi < \pi_0 / 2$ and $L_{sed} < 25$ the following should hold:
\begin{equation}
\frac{\pi_0}{\sigmapi} < 10 \sqrt{2}\label{eq:lsed_ok}
\end{equation}

If \autoref{eq:lsed_ok} holds, then the minimum value of $S$ is:

\begin{align}
S_{min} = (0.2\ln(10))^{2} \frac{\pi_0^2}{8 \sigmapi^2} < 25 (0.2\ln(10))^{2} \approx 5.3
\end{align}
For a typical values of $\sigma_{m_\lambda} \leq 0.05$, $S_{min} \ll \sum_\lambda \frac{1}{\sigma_{m_\lambda}^2}$, thus the contribution of $S_{min}$ to $\dLdmu$ can be neglected.

\subsection{Proof of the Statement 2}
Here, we want to prove that we can safely replace $\pi$ with $\pi_0$ in \autoref{eq:app_lsed_2nd_der}. To do that we want to show that:
\begin{equation}
\frac{\dLdmu(\pi)}{\dLdmu(\pi=\pi_0)} \simeq 1,
\end{equation}
or, equivalently, that \textbf{the fractional error $R$ of $\dLdmu(\pi)$ introduced by replacing $\pi$ with $\pi_0$ in \autoref{eq:app_lsed_2nd_der} is}:
\begin{equation}
R = \frac{\left|\dLdmu(\pi) - \dLdmu(\pi=\pi_0)\right|}{\dLdmu(\pi=\pi_0)} \ll 1, \label{eq:statement_2}
\end{equation}

Let us assume that $\pi = \pi_0 + k \sigmapi$, where $k$ should not be very large, so that $\frac{(\pi - \pi_0)^2}{2\sigmapi^2} = \frac{k^2}{2} < L_{sed} < 25$ -- otherwise $L_{sed}$ for a given model will be too large for this model to have a non-negligible contribution to PDFs.

Substituting $\pi = \pi_0 + k \sigmapi$ to Equations~\ref{eq:app_lsed_2nd_der} we get:
\begin{align}
\dLdmu &= (0.2\ln(10))^{2} \cdot \frac{(\pi_0 + k \sigmapi)(\pi_0 + 2 k \sigmapi)}{\sigmapi^2} +\sum_\lambda \frac{1}{\sigma_{m_\lambda}^2} = \nonumber \\
& (0.2\ln(10))^{2} \cdot \frac{(\pi_0^2 + 3 k \sigmapi \pi_0 + 2 k^2 \sigmapi^2)}{\sigmapi^2} +\sum_\lambda \frac{1}{\sigma_{m_\lambda}^2}.
\end{align}

Now, $R$ from the \autoref{eq:statement_2} will look like:
\begin{equation}
R = \frac{(0.2\ln(10))^{2} \cdot \frac{(3 k \sigmapi \pi_0 + 2 k^2 \sigmapi^2)}{\sigmapi^2}}{\dLdmu(\pi=\pi_0)} = \frac{(0.2\ln(10))^2 \cdot \left(\frac{3 k \pi_0}{\sigmapi} + 2 k^2\right)}
{(0.2\ln(10))^{2} \cdot \frac{\pi_0^2 }{\sigmapi^2} +\sum_\lambda \frac{1}{\sigma_{m_\lambda}^2}} \label{eq:statement_2_a}
\end{equation}

For simplicity, we designate $t = (0.2\ln(10))^{2}$ and $\frac{n}{\tilde{\sigma}_\lambda^2} = \sum_\lambda \frac{1}{\sigma_{m_\lambda}^2}$, where $n$ is the number of magnitudes used in $L_{sed}$ and $\tilde{\sigma}_\lambda$ is the average magnitude uncertainty. Using this, we can find the location of the maximum of $R$ as a function of $x=\frac{\pi_0}{\sigmapi}$ by solving the equation:
\begin{equation}
R'(x) = \left(\frac{t^2 (3 k x + 2 k ^2)}{t^2 x^2 + \frac{n}{\tilde{\sigma}_\lambda^2}}\right)' = 0
\end{equation}

Solving this equation for $x$ and substituting to \autoref{eq:statement_2_a} we arrive, after some algebra, to the following expression for the maximum value of $R(x)$:
\begin{equation}
R_{max}(k, \tilde{\sigma}_\lambda) = \frac{9}{2} \frac{k}{\sqrt{4 k^2 + \frac{9 n}{t^2 \tilde{\sigma}_\lambda^2 }}-2 k}\label{eq:rmax}
\end{equation}
This function indicates the fractional error that we bring into $H_{0,0}$ by substituting $\pi$ with $\pi_0$, as a function of $k$ and average photometric uncertainty $\tilde{\sigma}_\lambda$, maximised over all possible $x=\frac{\pi_0}{\sigmapi}$.

\begin{figure}
    \myimageTwo{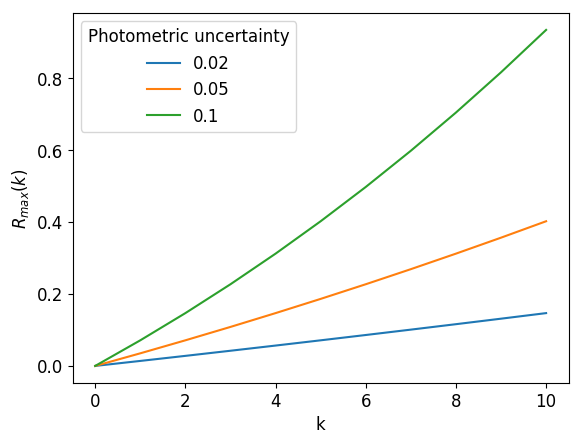}{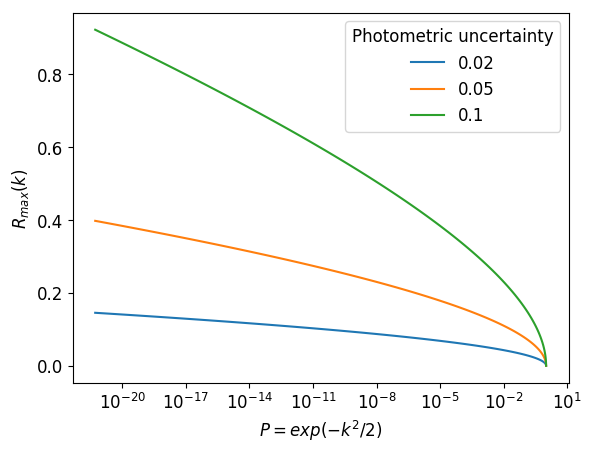}
    \caption{Fractional error of distance modulus uncertainty for a single model of a star, as a function of photometric uncertainty and offset $k$ (in units of parallax uncertainty) from the Gaia parallax $\pi_0$ (above) or model probability (below) }\label{fig:statement2}
\end{figure}

The function $R_{max}(k, \tilde{\sigma}_\lambda)$ is shown for three values of $\tilde{\sigma}_\lambda$ ($0^m.02$ for typical photometry, $0^m.05$ for bad photometry and $0^m.1$ for worst cases) in the top panel of \autoref{fig:statement2}. Bottom panel of \autoref{fig:statement2} shows the same data but now with chi-square probability values $P = e^{-k^2 / 2}$ used as x-axis.

In UniDAM we neglect all solutions for which the chi-squared probability of the best model is less than $0.1$. In the case when magnitudes in all photometric bands are available, this corresponds to $L_{sed} \approx 12$. Thus $\frac{k^2}{2} < 12$ and $k < 5$. For $k=5$ we have from \autoref{eq:rmax} that $R_{max}(k, \tilde{\sigma}_\lambda) < 0.5$.

This means that the error that we bring in by substituting $\pi$ with $\pi_0$ in \autoref{eq:app_lsed_2nd_der} is less than $50\%$ even for models that have a tiny contribution to the PDFs. Even though this $50\%$ might lead to a wrongly calculated uncertainties, it is very likely that for cases when both parallax and photometry have high uncertainties other effects, like the unknown systematics, will be more significant. 

\section{Additional figures}

\begin{figure*}
    \myimageTwo{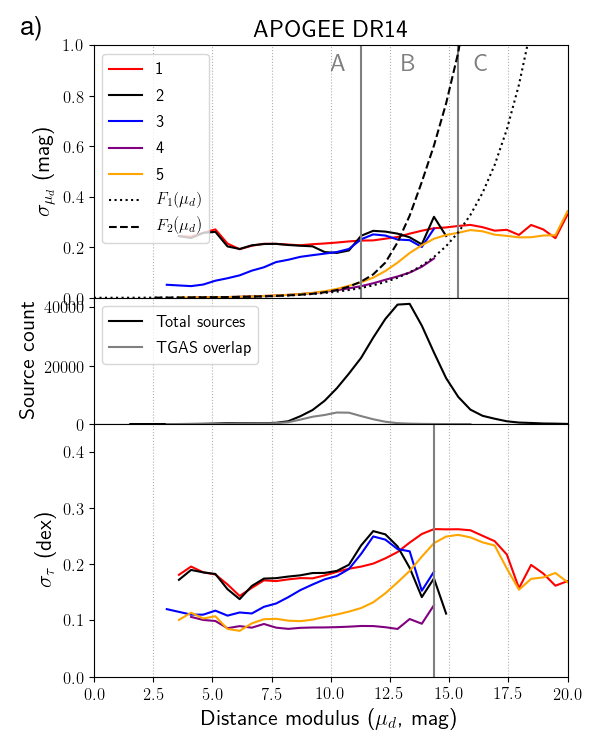}{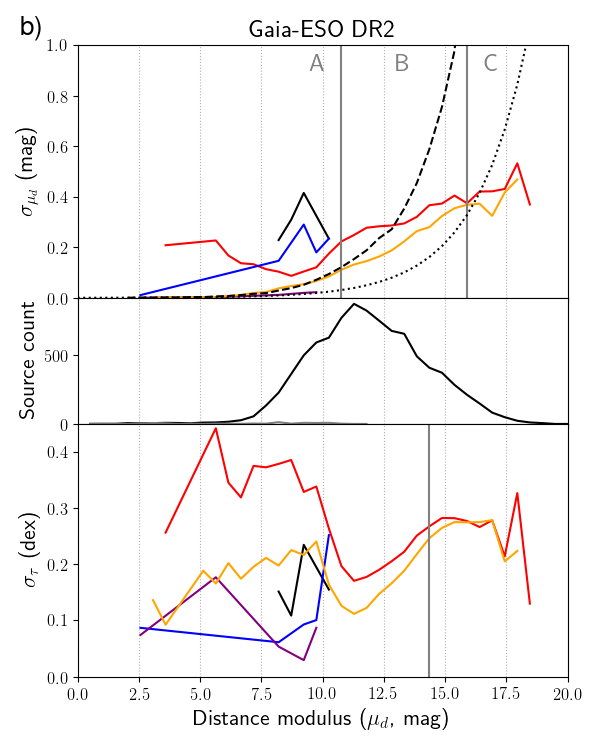}\par
    \myimageTwo{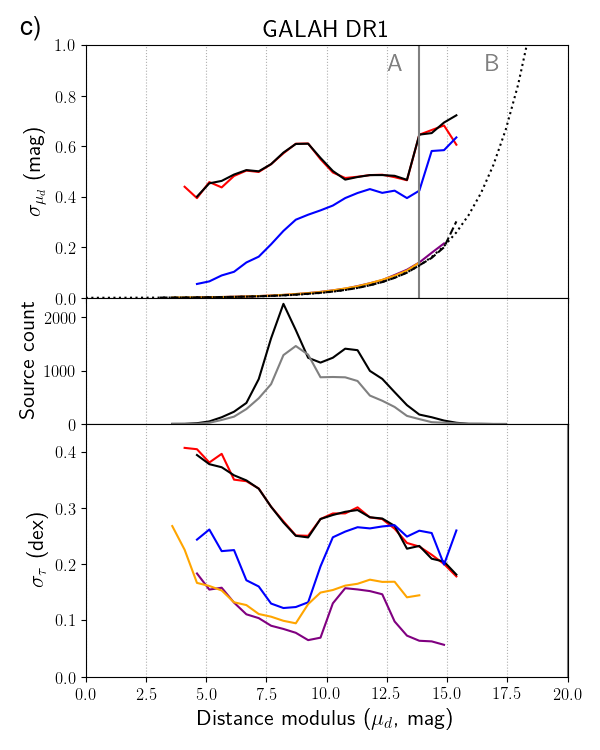}{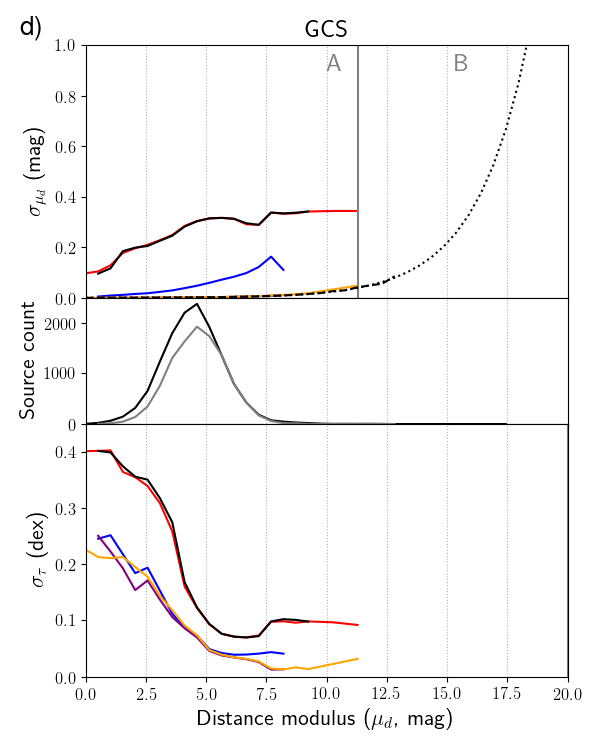}
    \caption{Same as \autoref{fig:uncertainties}}\label{fig:unc2}
\end{figure*}
\begin{figure*}\ContinuedFloat
    \myimageTwo{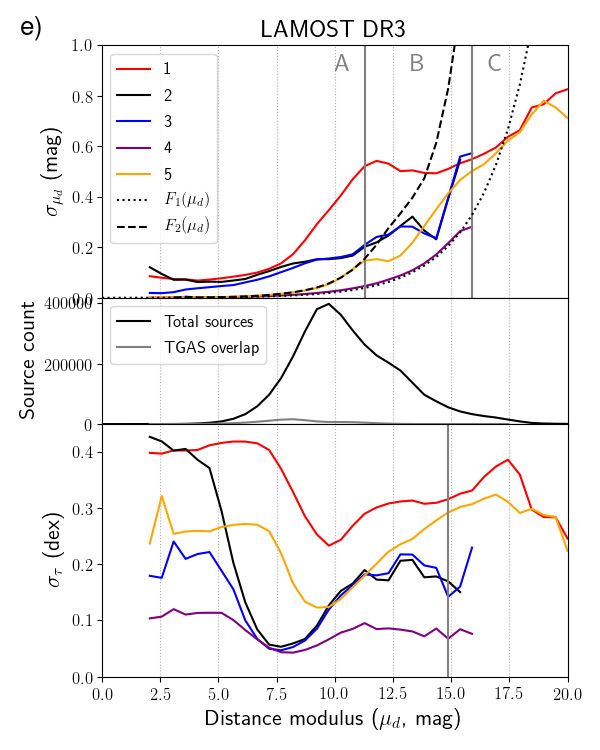}{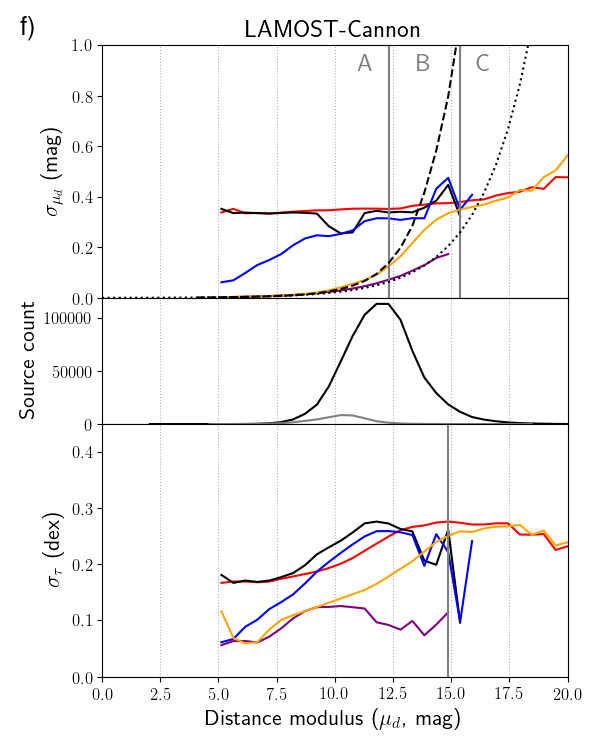}\par
    \myimageTwo{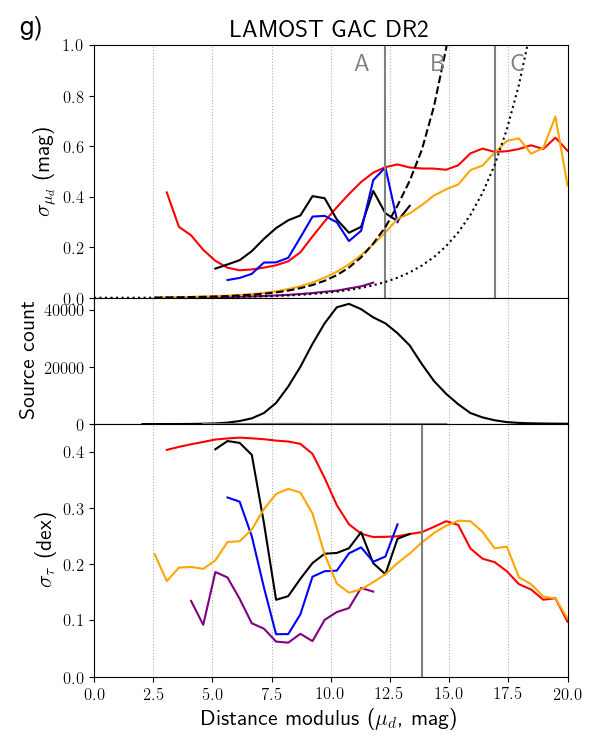}{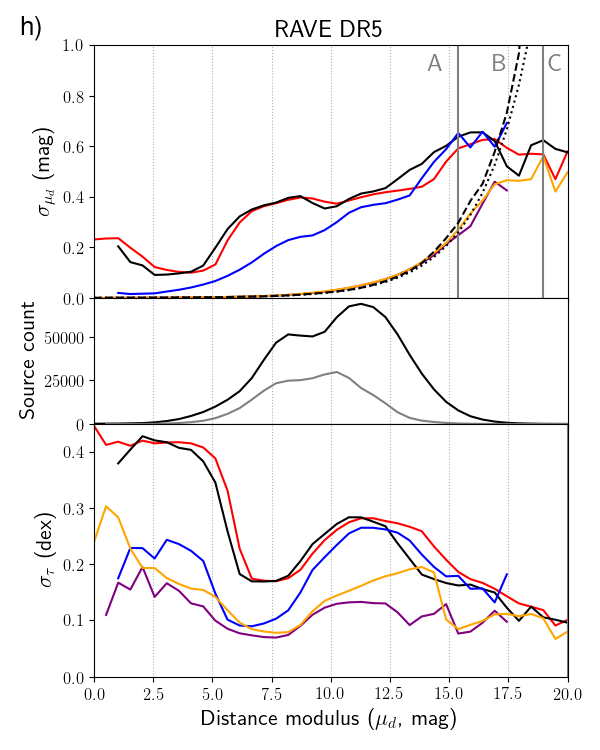}
    \caption{continued}\label{fig:unc3}
\end{figure*}
\begin{figure*}\ContinuedFloat
    \myimageTwo{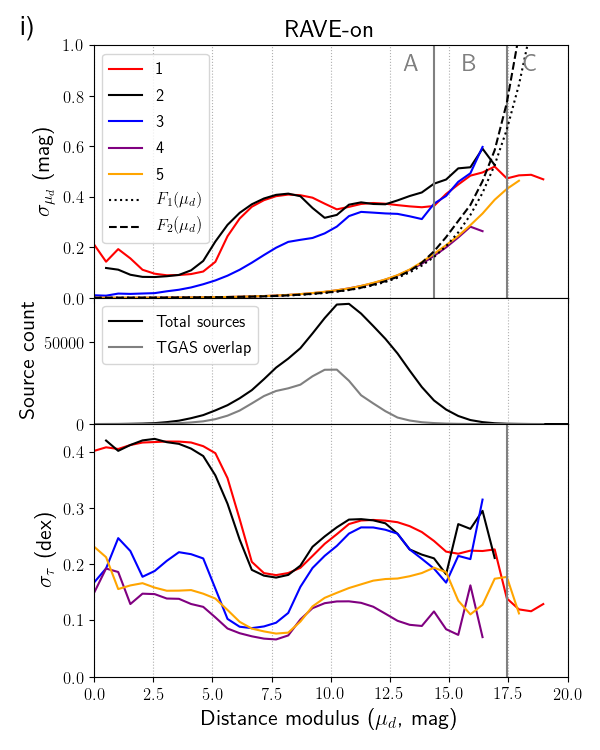}{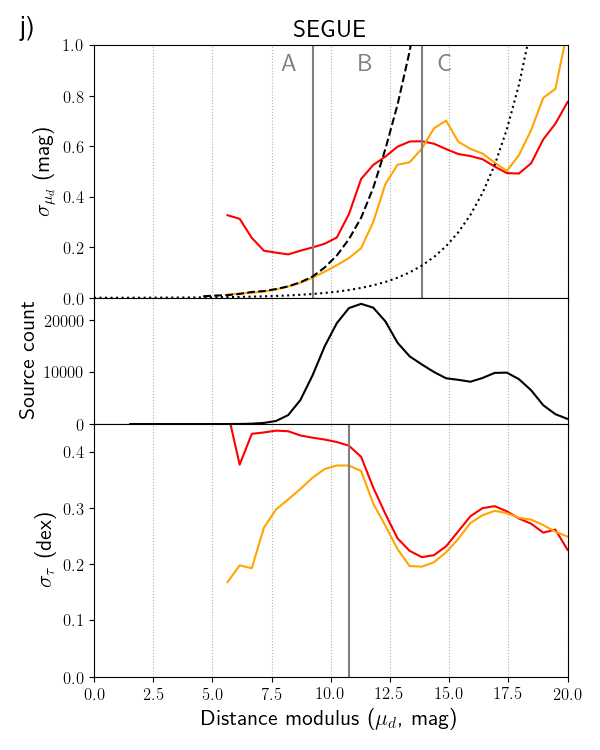}
    \myimages{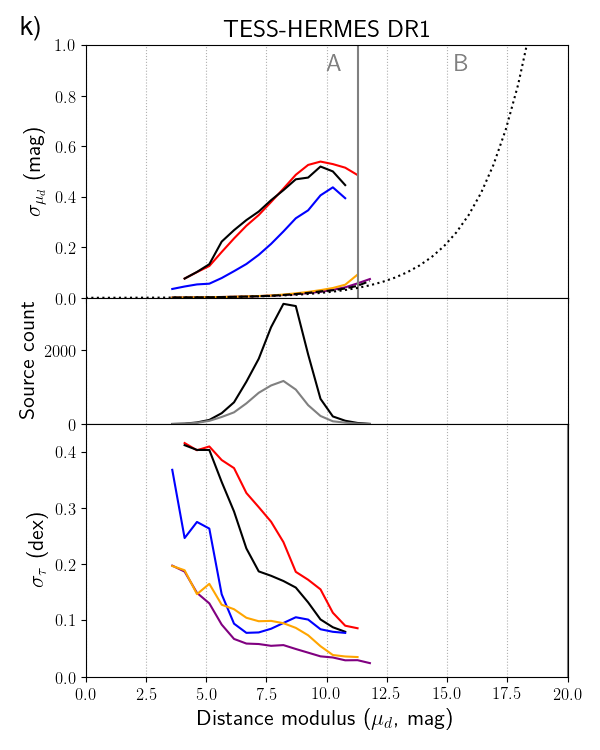}{0.45}
    \caption{continued. Note: for SEGUE survey (subplot j) there are no stars with TGAS parallaxes.}\label{fig:unc4}
\end{figure*}

\bibliographystyle{aa.bst}
\bibliography{sage_gaia}

\begin{thebibliography}{33}
\expandafter\ifx\csname natexlab\endcsname\relax\def\natexlab#1{#1}\fi

\bibitem[{{Abolfathi} {et~al.}(2017){Abolfathi}, {Aguado}, {Aguilar}, {Allende
  Prieto}, {Almeida}, {Tasnim Ananna}, {Anders}, {Anderson}, {Andrews},
  {Anguiano}, \& et~al.}]{2017arXiv170709322A}
{Abolfathi}, B., {Aguado}, D.~S., {Aguilar}, G., {et~al.} 2017, ArXiv e-prints
  [\eprint[arXiv]{1707.09322}]

\bibitem[{{Am{\^o}res} {et~al.}(2017){Am{\^o}res}, {Robin}, \&
  {Reyl{\'e}}}]{2017A&A...602A..67A}
{Am{\^o}res}, E.~B., {Robin}, A.~C., \& {Reyl{\'e}}, C. 2017, \aap, 602, A67

\bibitem[{{Astraatmadja} \& {Bailer-Jones}(2016)}]{2016ApJ...832..137A}
{Astraatmadja}, T.~L. \& {Bailer-Jones}, C.~A.~L. 2016, \apj, 832, 137

\bibitem[{{Astropy Collaboration} {et~al.}(2013){Astropy Collaboration},
  {Robitaille}, {Tollerud}, {Greenfield}, {Droettboom}, {Bray}, {Aldcroft},
  {Davis}, {Ginsburg}, {Price-Whelan}, {Kerzendorf}, {Conley}, {Crighton},
  {Barbary}, {Muna}, {Ferguson}, {Grollier}, {Parikh}, {Nair}, {Unther},
  {Deil}, {Woillez}, {Conseil}, {Kramer}, {Turner}, {Singer}, {Fox}, {Weaver},
  {Zabalza}, {Edwards}, {Azalee Bostroem}, {Burke}, {Casey}, {Crawford},
  {Dencheva}, {Ely}, {Jenness}, {Labrie}, {Lim}, {Pierfederici}, {Pontzen},
  {Ptak}, {Refsdal}, {Servillat}, \& {Streicher}}]{2013A&A...558A..33A}
{Astropy Collaboration}, {Robitaille}, T.~P., {Tollerud}, E.~J., {et~al.} 2013,
  \aap, 558, A33

\bibitem[{{Bressan} {et~al.}(2012){Bressan}, {Marigo}, {Girardi}, {Salasnich},
  {Dal Cero}, {Rubele}, \& {Nanni}}]{PARSEC}
{Bressan}, A., {Marigo}, P., {Girardi}, L., {et~al.} 2012, \mnras, 427, 127

\bibitem[{{Casagrande} {et~al.}(2011){Casagrande}, {Sch{\"o}nrich}, {Asplund},
  {Cassisi}, {Ram{\'{\i}}rez}, {Mel{\'e}ndez}, {Bensby}, \& {Feltzing}}]{GCS}
{Casagrande}, L., {Sch{\"o}nrich}, R., {Asplund}, M., {et~al.} 2011, \aap, 530,
  A138

\bibitem[{{Casey} {et~al.}(2017){Casey}, {Hawkins}, {Hogg}, {Ness}, {Rix},
  {Kordopatis}, {Kunder}, {Steinmetz}, {Koposov}, {Enke}, {Sanders}, {Gilmore},
  {Zwitter}, {Freeman}, {Casagrande}, {Matijevi{\v c}}, {Seabroke},
  {Bienaym{\'e}}, {Bland-Hawthorn}, {Gibson}, {Grebel}, {Helmi}, {Munari},
  {Navarro}, {Reid}, {Siebert}, \& {Wyse}}]{RAVE_ON}
{Casey}, A.~R., {Hawkins}, K., {Hogg}, D.~W., {et~al.} 2017, \apj, 840, 59

\bibitem[{{Dalton} {et~al.}(2014){Dalton}, {Trager}, {Abrams}, {Bonifacio},
  {L{\'o}pez Aguerri}, {Middleton}, {Benn}, {Dee}, {Say{\`e}de}, {Lewis},
  {Pragt}, {Pico}, {Walton}, {Rey}, {Allende Prieto}, {Pe{\~n}ate}, {Lhome},
  {Ag{\'o}cs}, {Alonso}, {Terrett}, {Brock}, {Gilbert}, {Ridings}, {Guinouard},
  {Verheijen}, {Tosh}, {Rogers}, {Steele}, {Stuik}, {Tromp}, {Jasko}, {Kragt},
  {Lesman}, {Mottram}, {Bates}, {Gribbin}, {Rodriguez}, {Delgado}, {Martin},
  {Cano}, {Navarro}, {Irwin}, {Lewis}, {Gonzalez Solares}, {O'Mahony},
  {Bianco}, {Zurita}, {ter Horst}, {Molinari}, {Lodi}, {Guerra}, {Vallenari},
  \& {Baruffolo}}]{WEAVE}
{Dalton}, G., {Trager}, S., {Abrams}, D.~C., {et~al.} 2014, in \procspie, Vol.
  9147, Ground-based and Airborne Instrumentation for Astronomy V, 91470L

\bibitem[{{de Jong} {et~al.}(2016){de Jong}, {Barden}, {Bellido-Tirado},
  {Brynnel}, {Frey}, {Giannone}, {Haynes}, {Johl}, {Phillips}, {Schnurr},
  {Walcher}, {Winkler}, {Ansorge}, {Feltzing}, {McMahon}, {Baker}, {Caillier},
  {Dwelly}, {Gaessler}, {Iwert}, {Mandel}, {Piskunov}, {Pragt}, {Walton},
  {Bensby}, {Bergemann}, {Chiappini}, {Christlieb}, {Cioni}, {Driver},
  {Finoguenov}, {Helmi}, {Irwin}, {Kitaura}, {Kneib}, {Liske}, {Merloni},
  {Minchev}, {Richard}, \& {Starkenburg}}]{4MOST}
{de Jong}, R.~S., {Barden}, S.~C., {Bellido-Tirado}, O., {et~al.} 2016, in
  \procspie, Vol. 9908, Ground-based and Airborne Instrumentation for Astronomy
  VI, 99081O

\bibitem[{{Gilmore} {et~al.}(2012){Gilmore}, {Randich}, {Asplund}, {Binney},
  {Bonifacio}, {Drew}, {Feltzing}, {Ferguson}, {Jeffries}, {Micela}, \&
  et~al.}]{GAIA_ESO}
{Gilmore}, G., {Randich}, S., {Asplund}, M., {et~al.} 2012, The Messenger, 147,
  25

\bibitem[{{Green} {et~al.}(2018){Green}, {Schlafly}, {Finkbeiner}, {Rix},
  {Martin}, {Burgett}, {Draper}, {Flewelling}, {Hodapp}, {Kaiser}, {Kudritzki},
  {Magnier}, {Metcalfe}, {Tonry}, {Wainscoat}, \&
  {Waters}}]{2018arXiv180103555G}
{Green}, G.~M., {Schlafly}, E.~F., {Finkbeiner}, D., {et~al.} 2018, ArXiv
  e-prints [\eprint[arXiv]{1801.03555}]

\bibitem[{{Ho} {et~al.}(2016){Ho}, {Ness}, {Hogg}, {Rix}, {Liu}, {Yang},
  {Zhang}, {Hou}, \& {Wang}}]{LAMOST_Cannon}
{Ho}, A.~Y.~Q., {Ness}, M.~K., {Hogg}, D.~W., {et~al.} 2016, ArXiv e-prints
  [\eprint[arXiv]{1602.00303}]

\bibitem[{Hunter(2007)}]{Hunter:2007}
Hunter, J.~D. 2007, Computing In Science \& Engineering, 9, 90

\bibitem[{Jones {et~al.}(2001--)Jones, Oliphant, Peterson, {et~al.}}]{SciPy}
Jones, E., Oliphant, T., Peterson, P., {et~al.} 2001--, {SciPy}: Open source
  scientific tools for {Python}, [Online; accessed <today>]

\bibitem[{{Katz} \& {Brown}(2017)}]{2017arXiv171010816K}
{Katz}, D. \& {Brown}, A.~G.~A. 2017, ArXiv e-prints
  [\eprint[arXiv]{1710.10816}]

\bibitem[{{Kovalevsky}(1998)}]{1998A&A...340L..35K}
{Kovalevsky}, J. 1998, \aap, 340, L35

\bibitem[{{Kunder} {et~al.}(2016){Kunder}, {Kordopatis}, {Steinmetz},
  {Zwitter}, {McMillan}, {Casagrande}, {Enke}, {Wojno}, {Valentini},
  {Chiappini}, {Matijevic}, {Siviero}, {de Laverny}, {Recio-Blanco}, {Bijaoui},
  {Wyse}, {Binney}, {Grebel}, {Helmi}, {Jofre}, {Gilmore}, {Siebert}, {Famaey},
  {Bienayme}, {Gibson}, {Freeman}, {Navarro}, {Munari}, {Seabroke}, {Anguiano
  Jimenez}, {Reid}, {Bland-Hawthorn}, {Watson}, {Gerhard}, {Davies},
  {Elsworth}, {Lund}, {Miglio}, {Chaplin}, \& {Mosser}}]{RAVE_DR5}
{Kunder}, A., {Kordopatis}, G., {Steinmetz}, M., {et~al.} 2016, ArXiv e-prints
  [\eprint[arXiv]{1609.03210}]

\bibitem[{{Lindegren} {et~al.}(2016){Lindegren}, {Lammers}, {Bastian},
  {Hernández}, {Klioner}, {Hobbs}, {Bombrun}, {Michalik}, {Ramos-Lerate},
  {Butkevich}, {Comoretto}, {Joliet}, {Holl}, {Hutton}, {Parsons},
  {Steidelmüller}, {Abbas}, {Altmann}, {Andrei}, {Anton}, {Bach}, {Barache},
  {Becciani}, {Berthier}, {Bianchi}, {Biermann}, {Bouquillon}, {Bourda},
  {Brüsemeister}, {Bucciarelli}, {Busonero}, {Carlucci}, {Castañeda},
  {Charlot}, {Clotet}, {Crosta}, {Davidson}, {de Felice}, {Drimmel},
  {Fabricius}, {Fienga}, {Figueras}, {Fraile}, {Gai}, {Garralda}, {Geyer},
  {González-Vidal}, {Guerra}, {Hambly}, {Hauser}, {Jordan}, {Lattanzi},
  {Lenhardt}, {Liao}, {Löffler}, {McMillan}, {Mignard}, {Mora}, {Morbidelli},
  {Portell}, {Riva}, {Sarasso}, {Serraller}, {Siddiqui}, {Smart}, {Spagna},
  {Stampa}, {Steele}, {Taris}, {Torra}, {van Reeven}, {Vecchiato}, {Zschocke},
  {de Bruijne}, {Gracia}, {Raison}, {Lister}, {Marchant}, {Messineo}, {Soffel},
  {Osorio}, {de Torres}, \& {O'Mullane}}]{2016A&A...595A...4L}
{Lindegren}, L., {Lammers}, U., {Bastian}, U., {et~al.} 2016, \aap, 595, A4

\bibitem[{{Luo} {et~al.}(2015){Luo}, {Zhao}, {Zhao}, {Deng}, {Liu}, {Jing},
  {Wang}, {Zhang}, {Shi}, {Cui}, {Chu}, {Li}, {Bai}, {Wu}, {Cai}, {Cao}, {Cao},
  {Carlin}, {Chen}, {Chen}, {Chen}, {Chen}, {Chen}, {Chen}, {Chen},
  {Christlieb}, {Chu}, {Cui}, {Dong}, {Du}, {Fan}, {Feng}, {Fu}, {Gao}, {Gong},
  {Gu}, {Guo}, {Han}, {He}, {Hou}, {Hou}, {Hou}, {Hu}, {Hu}, {Hu}, {Huo},
  {Jia}, {Jiang}, {Jiang}, {Jiang}, {Jin}, {Kong}, {Kong}, {Lei}, {Li}, {Li},
  {Li}, {Li}, {Li}, {Li}, {Li}, {Li}, {Li}, {Li}, {Li}, {Li}, {Liang}, {Lin},
  {Liu}, {Liu}, {Liu}, {Liu}, {Lu}, {Luo}, {Mao}, {Newberg}, {Ni}, {Qi}, {Qi},
  {Shen}, {Shi}, {Song}, {Song}, {Su}, {Su}, {Tang}, {Tao}, {Tian}, {Wang},
  {Wang}, {Wang}, {Wang}, {Wang}, {Wang}, {Wang}, {Wang}, {Wang}, {Wang},
  {Wang}, {Wang}, {Wang}, {Wang}, {Wang}, {Wang}, {Wang}, {Wang}, {Wang},
  {Wang}, {Wei}, {Wei}, {Wu}, {Wu}, {Wu}, {Wu}, {Xing}, {Xu}, {Xu}, {Xu},
  {Yan}, {Yang}, {Yang}, {Yang}, {Yang}, {Yao}, {Yu}, {Yuan}, {Yuan}, {Yuan},
  {Yuan}, {Zhai}, {Zhang}, {Zhang}, {Zhang}, {Zhang}, {Zhang}, {Zhang},
  {Zhang}, {Zhang}, {Zhao}, {Zhou}, {Zhou}, {Zhu}, {Zhu}, {Zou}, \&
  {Zuo}}]{LAMOST}
{Luo}, A.-L., {Zhao}, Y.-H., {Zhao}, G., {et~al.} 2015, Research in Astronomy
  and Astrophysics, 15, 1095

\bibitem[{{Mackereth} {et~al.}(2017){Mackereth}, {Bovy}, {Schiavon},
  {Zasowski}, {Cunha}, {Frinchaboy}, {Garcia Perez}, {Hayden}, {Holtzman},
  {Majewski}, {Meszaros}, {Nidever}, {Pinsonneault}, \&
  {Shetrone}}]{2017arXiv170600018M}
{Mackereth}, J.~T., {Bovy}, J., {Schiavon}, R.~P., {et~al.} 2017, ArXiv
  e-prints [\eprint[arXiv]{1706.00018}]

\bibitem[{{Martell} {et~al.}(2016){Martell}, {Sharma}, {Buder}, {Duong},
  {Schlesinger}, {Simpson}, {Lind}, {Ness}, {Marshall}, {Asplund},
  {Bland-Hawthorn}, {Casey}, {De Silva}, {Freeman}, {Kos}, {Lin}, {Zucker},
  {Zwitter}, {Anguiano}, {Bacigalupo}, {Carollo}, {Casagrande}, {Da Costa},
  {Horner}, {Huber}, {Hyde}, {Kafle}, {Lewis}, {Nataf}, {Stello}, {Tinney},
  {Watson}, \& {Wittenmyer}}]{GALAH}
{Martell}, S., {Sharma}, S., {Buder}, S., {et~al.} 2016, ArXiv e-prints
  [\eprint[arXiv]{1609.02822}]

\bibitem[{{Martig} {et~al.}(2016){Martig}, {Minchev}, {Ness}, {Fouesneau}, \&
  {Rix}}]{2016ApJ...831..139M}
{Martig}, M., {Minchev}, I., {Ness}, M., {Fouesneau}, M., \& {Rix}, H.-W. 2016,
  \apj, 831, 139

\bibitem[{{McMillan} {et~al.}(2017){McMillan}, {Kordopatis}, {Kunder},
  {Binney}, {Wojno}, {Zwitter}, {Steinmetz}, {Bland-Hawthorn}, {Gibson},
  {Gilmore}, {Grebel}, {Helmi}, {Munari}, {Navarro}, {Parker}, {Seabroke}, \&
  {Wyse}}]{2017arXiv170704554M}
{McMillan}, P.~J., {Kordopatis}, G., {Kunder}, A., {et~al.} 2017, ArXiv
  e-prints [\eprint[arXiv]{1707.04554}]

\bibitem[{{Michalik} {et~al.}(2015){Michalik}, {Lindegren}, \&
  {Hobbs}}]{2015A&A...574A.115M}
{Michalik}, D., {Lindegren}, L., \& {Hobbs}, D. 2015, \aap, 574, A115

\bibitem[{{Mints} \& {Hekker}(2017)}]{Paper1}
{Mints}, A. \& {Hekker}, S. 2017, ArXiv e-prints [\eprint[arXiv]{1705.00963}]

\bibitem[{{Queiroz} {et~al.}(2018){Queiroz}, {Anders}, {Santiago}, {Chiappini},
  {Steinmetz}, {Ponte}, {Stassun}, {da Costa}, {Maia}, {Crestani}, {Beers},
  {Fern{\'a}ndez-Trincado}, {Garc{\'{\i}}a-Hern{\'a}ndez}, {Roman-Lopes}, \&
  {Zamora}}]{2018MNRAS.tmp..326Q}
{Queiroz}, A.~B.~A., {Anders}, F., {Santiago}, B.~X., {et~al.} 2018, \mnras
  [\eprint[arXiv]{1710.09970}]

\bibitem[{{Recio-Blanco} {et~al.}(2016){Recio-Blanco}, {de Laverny}, {Allende
  Prieto}, {Fustes}, {Manteiga}, {Arcay}, {Bijaoui}, {Dafonte}, {Ordenovic}, \&
  {Ordo{\~n}ez Blanco}}]{2016A&A...585A..93R}
{Recio-Blanco}, A., {de Laverny}, P., {Allende Prieto}, C., {et~al.} 2016,
  \aap, 585, A93

\bibitem[{{Schlegel} {et~al.}(1998){Schlegel}, {Finkbeiner}, \&
  {Davis}}]{1998ApJ...500..525S}
{Schlegel}, D.~J., {Finkbeiner}, D.~P., \& {Davis}, M. 1998, \apj, 500,
  525–553

\bibitem[{{Sharma} {et~al.}(2017){Sharma}, {Stello}, {Buder}, {Kos},
  {Bland-Hawthorn}, {Asplund}, {Duong}, {Lin}, {Lind}, {Ness}, {Huber},
  {Zwitter}, {Hon}, {Kafle}, {Khanna}, {Saddon}, {Anguiano}, {Casey},
  {Freeman}, {Martell}, {De Silva}, {Simpson}, {Wittenmyer}, \&
  {Zucker}}]{TESS_HERMES}
{Sharma}, S., {Stello}, D., {Buder}, S., {et~al.} 2017, ArXiv e-prints
  [\eprint[arXiv]{1707.05753}]

\bibitem[{{Soderblom}(2010)}]{2010ARA&A..48..581S}
{Soderblom}, D.~R. 2010, \araa, 48, 581

\bibitem[{{Taylor}(2005)}]{2005ASPC..347...29T}
{Taylor}, M.~B. 2005, in Astronomical Society of the Pacific Conference Series,
  Vol. 347, Astronomical Data Analysis Software and Systems XIV, ed.
  P.~{Shopbell}, M.~{Britton}, \& R.~{Ebert}, 29

\bibitem[{{Xiang} {et~al.}(2017){Xiang}, {Liu}, {Yuan}, {Huo}, {Huang}, {Wang},
  {Chen}, {Ren}, {Zhang}, {Tian}, {Yang}, {Shi}, {Zhao}, {Li}, {Zhao}, {Cui},
  {Li}, {Hou}, {Zhang}, {Zhang}, {Wang}, {Wu}, {Cao}, {Yan}, {Yan}, {Luo},
  {Zhang}, {Bai}, {Yuan}, {Dong}, {Lei}, \& {Li}}]{2017arXiv170105409X}
{Xiang}, M., {Liu}, X., {Yuan}, H., {et~al.} 2017, ArXiv e-prints
  [\eprint[arXiv]{1701.05409}]

\bibitem[{{Yanny} {et~al.}(2009){Yanny}, {Rockosi}, {Newberg}, {Knapp},
  {Adelman-McCarthy}, {Alcorn}, {Allam}, {Allende Prieto}, {An}, {Anderson},
  {Anderson}, {Bailer-Jones}, {Bastian}, {Beers}, {Bell}, {Belokurov},
  {Bizyaev}, {Blythe}, {Bochanski}, {Boroski}, {Brinchmann}, {Brinkmann},
  {Brewington}, {Carey}, {Cudworth}, {Evans}, {Evans}, {Gates}, {G{\"a}nsicke},
  {Gillespie}, {Gilmore}, {Nebot Gomez-Moran}, {Grebel}, {Greenwell}, {Gunn},
  {Jordan}, {Jordan}, {Harding}, {Harris}, {Hendry}, {Holder}, {Ivans},
  {Ivezi{\v c}}, {Jester}, {Johnson}, {Kent}, {Kleinman}, {Kniazev},
  {Krzesinski}, {Kron}, {Kuropatkin}, {Lebedeva}, {Lee}, {French Leger},
  {L{\'e}pine}, {Levine}, {Lin}, {Long}, {Loomis}, {Lupton}, {Malanushenko},
  {Malanushenko}, {Margon}, {Martinez-Delgado}, {McGehee}, {Monet}, {Morrison},
  {Munn}, {Neilsen}, {Nitta}, {Norris}, {Oravetz}, {Owen}, {Padmanabhan},
  {Pan}, {Peterson}, {Pier}, {Platson}, {Re Fiorentin}, {Richards}, {Rix},
  {Schlegel}, {Schneider}, {Schreiber}, {Schwope}, {Sibley}, {Simmons},
  {Snedden}, {Allyn Smith}, {Stark}, {Stauffer}, {Steinmetz}, {Stoughton},
  {SubbaRao}, {Szalay}, {Szkody}, {Thakar}, {Sivarani}, {Tucker}, {Uomoto},
  {Vanden Berk}, {Vidrih}, {Wadadekar}, {Watters}, {Wilhelm}, {Wyse}, {Yarger},
  \& {Zucker}}]{SEGUE}
{Yanny}, B., {Rockosi}, C., {Newberg}, H.~J., {et~al.} 2009, \aj, 137, 4377

\end{thebibliography}

\end{document}